\documentclass[printer]{aa}
\usepackage{graphicx}
\usepackage{txfonts}
\usepackage{natbib}
\usepackage[bookmarks=true,bookmarksopen=true,pdfstartview = FitV,pdfpagelayout = SinglePage,colorlinks = true,linkcolor=blue,citecolor=blue,urlcolor = blue,pdfborder = {0 0 0}]{hyperref} 
\bibpunct{(}{)}{;}{a}{}{,}
\let\oldbibliography\thebibliography
\renewcommand{\thebibliography}[1]{
  \oldbibliography{#1}
  \setlength{\itemsep}{0pt}
}
\newcommand{\dd}{\mathrm{d}}
\newcommand{\sgra}{Sgr~A*}
\usepackage{pdfpages}
\begin{document}

\title{Study of the X-ray activity of \object{Sgr A*} \\ during the 2011 XMM-Newton campaign}
\author{Enmanuelle Mossoux \inst{\ref{inst1}}
\and Nicolas Grosso \inst{\ref{inst1}}
\and Frédéric H. Vincent \inst{\ref{inst2}}
\and Delphine Porquet \inst{\ref{inst1}}}
\institute{Observatoire Astronomique de Strasbourg, Université de Strasbourg, CNRS, UMR 7550, 11 rue de l’Université, F-67000 Strasbourg, France. \label{inst1}
\and Nicolaus Copernicus Astronomical Center, ul. Bartycka 18, 00-716 Warszawa, Poland.\label{inst2}}
\date{Received July 25, 2014 / Accepted  September 19, 2014}

\abstract
{At the dynamical center of the Milky Way, there is the closest supermassive black hole: \sgra{}.
Its non-flaring luminosity is several orders of magnitude lower than the Eddington luminosity, but flares can be observed in the infrared and X-rays.
This flaring activity can help us to understand radiation mechanisms from \sgra{}.} 
{Our aim is to investigate the X-ray flaring activity of \sgra{} and to constrain the physical properties of the X-ray flares and their origin.}
{In March and April 2011, we observed \sgra{} with XMM-Newton with a total exposure of $\approx 226\ \mathrm{ks}$ in coordination with the 1.3~mm Very-Long-Baseline Interferometry array.
We performed timing analysis of the X-ray emission from \sgra{} using a Bayesian-blocks algorithm to detect X-ray flares observed with XMM-Newton.
Furthermore, we computed X-ray smoothed light curves observed in this campaign in order to have better accuracy on the position and the amplitude of the flares.}
{We detected two X-ray flares on March 30 and April 3, 2011, which for comparison have a peak detection level of 6.8 and 5.9 $\sigma$ in the XMM-Newton/EPIC (pn+MOS1+MOS2) light curve in the 2$-$10~keV energy range with a 300 s bin.
The former is characterized by two sub-flares: the first one is very short ($\sim 458$ s) with a peak luminosity of $L\mathrm{^{unabs}_{2-10~keV}} \sim 9.4 \times 10^{34}\ \mathrm{erg\ s^{-1}}$, whereas the second one is longer ($\sim 1542$ s) with a lower peak luminosity ($L\mathrm{^{unabs}_{2-10~keV}} \sim 6.8 \times 10^{34}\ \mathrm{erg\ s^{-1}}$).
The comparison with the sample of X-ray flares detected during the 2012 \textit{Chandra XVP} campaign favors the hypothesis that the 2011 March 30 flare is a single flare rather than two distinct subflares.
We model the light curve of this flare with the gravitational lensing of a simple hotspot-like structure, but we cannot satisfactorily reproduce the large decay of the light curve between the two subflares with this model.
From magnetic energy heating during the rise phase of the first subflare and assuming an X-ray photons production efficiency of 1 and a magnetic field of 100 G at $2\ r_{\mathrm{g}}$, we derive an upper limit to the radial distance of the first subflare of $100^{+19}_{-29}\ r_{\mathrm{g}}$.
We use the decay phase of the first subflare to estimate a lower limit to the radial distance of $4 \ r_{\mathrm{g}}$ from synchrotron cooling in the infrared.}
{The X-ray emitting region of the first subflare is located at a radial position of $4-100^{+19}_{-29}$ and has a corresponding radius of $1.8- 2.87 \pm 0.01$ in $r_{\mathrm{g}}$ unit for a magnetic field of 100 G at $2\ r_{\mathrm{g}}$.}


\titlerunning{The 2011 XMM-Newton campaign of \sgra}
\maketitle

\section{Introduction} 
Our Galaxy hosts \sgra\  at its dynamical center. It is the closest supermassive black hole (SMBH) at a distance of about 8 kpc \citep{genzel10,falcke13}.
\sgra{} has a mass, $M_\mathrm{BH}$, of about $4 \times 10^6\ \mathrm{\textit{M}_{\sun}}$, which was determined thanks to the measurements of star motions \citep{schodel02,ghez08,gillessen09}.
The Galactic center SMBH is usually in a steady state, emitting predominately at radio to submillimeter wavelengths.
Its bolometric luminosity is about $10^{36}\ \mathrm{erg}\ \mathrm{s^{-1}}$ \citep{yuan03}, which corresponds to $\approx 2 \times 10^{-9}\ \mathrm{\textit{L}_{Edd}}$ with $\mathrm{\textit{L}_{Edd}}=3.3 \times 10^4(\mathrm{\textit{M}_\mathrm{BH}/\textit{M}_{\sun}})\mathrm{\textit{L}_{\sun}}$.
To explain this low luminosity, researchers have developed various mass-accretion flow models, such as the advection-dominated accretion flows (ADAF; \citealt{narayan98}) and jet-disk models like the ejection of magnetized plasma \citep{falcke93}.

\citet{wang13} have recently inferred the temperature and density profile of the X-ray emitting gas around \sgra with the help of deep Chandra observations.
They have shown that $\le 1$\% of the gas initially captured by the SMBH at the Bondi radius reaches the innermost region around \sgra; i.e, $\ge 99$\% of the gas is ejected, which is consistent with the predictions of radiatively inefficient accretion flow (RIAF) models.
Therefore, \sgra{} is the ideal astronomical target for investigating the physics of mass accretion and ejection onto SMBH in the regime of a low mass-accretion rate, a state where they are supposed to spend most of their lifetime \citep{ho08}. 
This physical understanding could then be extended to the normal galaxies that dominate the population of galaxies in the local Universe.

The detections of flares from \sgra{} (first discovered in X-rays; \citealt{baganoff01}) have provided a valuable way to scrutinize accreting matter close to the event horizon. 
The X-ray flare frequency is 1.1 (1.0-1.3) flare per day with $\ \mathrm{\textit{L}_{2-8\ \mathrm{keV}}} \ge 10^{34}\ \mathrm{erg\ s^{-1}}$ \citep{neilsen13}, though episodes of higher X-ray flaring activity can also be observed \citep{porquet08,neilsen13}. 
The bulk of X-ray flares detected so far have faint-to-moderate amplitudes with factors of about 2 to 45 compared to the non-flaring luminosity ($\mathrm{\textit{L}_{2-8\ \mathrm{keV}}} \approx 3.6 \times 10^{33}\ \mathrm{erg\ s^{-1}}$; \citealt{baganoff03,neilsen13}), and three very bright flares (factors of 100-160 times the non-flaring luminosity) have been observed to share very similar spectral properties \citep{porquet03,porquet08,nowak12}.
The light curves of the X-ray flares can exhibit deep drops with short duration indicating that the X-ray emission comes from a region as compact as seven Schwarzschild radii ($\mathrm{\textit{R}_{S}}\equiv 2G\textit{M}_{BH}/c^2=1.2 \times 10^{12}\ \mathrm{cm}$ for \sgra, i.e., $\approx 0.6\ \mathrm{AU}$; \citealt{porquet03}). 

When near-infrared (NIR) and X-ray flares are detected simultaneously, their light curves have similar shapes, and there is no apparent delay ($< 3\ \mathrm{min}$) between the peaks of flare emission (e.g., \citealt{yusef-zadeh06,dodds-eden09,eckart12}). 
The current interpretation is that both X-ray/NIR flares come from a region close to the event horizon, while delayed sub-mm (e.g., $\approx100\ \mathrm{min}$; \citealt{marrone08}) and mm peaks (up to 5 hours; \citealt{yusef-zadeh09}) have been interpreted as the adiabatic cooling of an expanding relativistic plasma blob. 
While NIR flares are known to be due to synchrotron emission \citep{eisenhauer05,eckart06}, the X-ray flare emission mechanism has not been settled yet, with arguments for synchrotron \citep{dodds-eden09,barriere14}, inverse Compton \citep{yusef-zadeh12}, and synchrotron self-Compton \citep{eckart08} models.

We report here the results of our \sgra{} observation campaign performed with XMM-Newton from March 28 to April 5, 2011 in coordination with the 1.3~mm Very-Long-Baseline Interferometry array (VLBI).
In Sect. \ref{observation} we describe the XMM-Newton observations and data processing.
In Sect. \ref{timing} we present our timing analysis of \sgra{}.
In Sect. \ref{spectrum} we describe the spectral analysis of the two flares from \sgra{} detected during this 2011 campaign.
In Sect. \ref{discussion} we compare these flares with those detected in the 2012 Chandra XVP campaign.
We also try to model the first subflare with a simple hotspot model and estimate a lower and upper limit to the radial distance of this subflare.
Finally, in Sect. \ref{summary} we summarize our main results.

\section{XMM-Newton observations and data processing} 
\label{observation}
\subsection{Observation set-up}
 These X-ray observations of Sgr A* with XMM-Newton (AO-8, $5 \times 33$ ks; PI: D. Porquet) were designed to perform the first simultaneous observational campaign in X-rays and at 1.3~mm with the VLBI \citep{doeleman08}, in order to constrain the location X-ray flares. 
Five observing nights with the 1.3~mm VLBI were planned in 2011 between March 28 and April 5, using the weather forecast each day at noon for the final optimized scheduling (PI: S.\ Doeleman). 
The merged visibility window of the 1.3~mm VLBI array formed by {\sl the Atacama Pathfinder Experiment} (APEX) in Chile, {\sl the Submillimeter Telescope} (SMT) in Arizona, {\sl the Combined Array for Research in Millimeter-wave Astronomy} (CARMA) in California, and {\sl the Submillimeter Array} (SMA) in Hawaii is 10:45–15:45~UT.
Since X-ray flare peaks appear to occur before submillimeter peak \citep{marrone08} and since they can last up to three hours \citep{baganoff01}, our XMM-Newton observations started about three hours before the VLBI visibility window. We observed \sgra{} with XMM-Newton continuously from about 07:40~UT to about~16:00 Universal Time (UT), which is a duration of 30 ks.
The XMM-Newton visibility windows finally constrained the five following dates: 2011 March 28 and 30 and April 1, 3, and 5.
The 1.3~mm VLBI observations were obtained on 2011 March 29 and 31 and April~1 (simultaneous with XMM-Newton), 2, and 4; the results of these observations will be reported elsewhere.
Two complementary Chandra observations were obtained to extend the X-ray coverage on 2011 March~29 and March~31 from 10:29~UT to 15:29~UT (Cycle 12; PI: F.\ Baganoff), the former being simultaneous with VLBI. The results of these observations will be reported elsewhere.

The two XMM-Newton/EPIC MOS cameras \citep{turner01} and the XMM-Newton/EPIC pn camera \citep{strueder01} were operated in the full frame window mode with the medium filter. 
EPIC pn camera starts to observe after EPIC MOS cameras and stops before them.
The effective starting and end times of each observation are reported in Table \ref{table:1}.
These times are the time of the beginning and the stop of the observation with EPIC MOS cameras in the terrestrial time (TT) referential.
For this observation, the relation between terrestrial time and UT is $\mathrm{UT}=\mathrm{TT}-66.18\mathrm{s}$ (NASA's HEASARC Tool: xTime\footnote{The website of xTime is: \href{http://heasarc.gsfc.nasa.gov/cgi-bin/Tools/xTime/xTime.pl}{http://heasarc.gsfc.nasa.gov/cgi-bin/Tools\-/xTime/xTime.pl}}).

\begin{table}
\caption{XMM-Newton observation log for the Spring 2011 campaign.}
\centering
\tiny
\label{table:1}
\begin{tabular}{c c c c c}
\hline
\hline
Orbit & ObsID & Start Time\tablefootmark{a} & End Time\tablefootmark{a} & Duration \\
 & & (TT) & (TT) & (s) \\
\hline
 2069 & 0604300601 & Mar. 28, 07:54:14 & Mar. 28, 21:13:55 & 47981 \\
 2070 & 0604300701 & Mar. 30, 08:11:26 & Mar. 30, 21:14:28 & 46942 \\
 2071 & 0604300801 & Apr. 01, 08:23:50 & Apr. 01, 19:23:59 & 39609 \\
 2072 & 0604300901 & Apr. 03, 07:56:23 & Apr. 03, 19:21:36 & 41113 \\
 2073 & 0604301001 & Apr. 05, 07:13:49 & Apr. 05, 21:11:49 & 50280 \\
\hline
\end{tabular}
\tablefoot{
\tablefoottext{a}
{Start and end times of the EPIC MOS camera observations in terrestrial time (TT) referential.}
}
\normalsize
\end{table}

\subsection{Data processing}
We observed \sgra{} five times with XMM-Newton in early 2011 for a total effective exposure of $\approx 226\ \mathrm{ks}$.
We use the version 13.5 of the Science Analysis Software (SAS) package for the data reduction and analysis, with the latest release of the current calibration files (CCF; as of 04/04/2014).
The MOS and pn event lists were produced using the SAS tasks \texttt{emchain} and \texttt{epchain}, respectively. 
The full detector light curves in the 2$-$10 keV energy range computed by these tasks reveal that the observation was only slightly affected by weak soft proton flares.
The count rate of these soft protons was high only during the last four, three, one, and four hours of the 1$^{\mathrm{st}}$, 2$^{\mathrm{nd}}$, 3$^{\mathrm{rd}}$, and 4$^{\mathrm{th}}$ observations, respectively.

We concentrate on analysis of the central point source, \sgra and, in particular, on the search for the variability of its X-ray emission.
To do this, we define the source+background region as a $10\arcsec$-radius disk around the VLBI radio position of \sgra: RA(J2000)=$17^{\mathrm{h}}45^{\mathrm{m}}40\fs{}0409$, DEC(J2000)=$-29^{\circ} 00\arcmin 28\farcs{}118$ \citep{reid99}.
We do not register the EPIC coordinates again because the absolute astrometry for the EPIC cameras is about $1\farcs{}2$ \citep{EPIC_calibration_status_document}.
To create the light curves, we selected the events for MOS and pn with \texttt{PATTERN$\leq 12$} and \texttt{\#XMMEA\_SM}, and \texttt{PATTERN$\leq 4$} and \texttt{FLAG==0}, respectively.
The contribution of the background proton flares was estimated using a $\approx 3\arcmin \times 3\arcmin$ area with a low level of X-ray extended emission, located on the same CCD at $\approx 4\arcmin$ -north of \sgra, where the X-ray emission of point sources were subtracted.
This data reduction is the same as in \citet{porquet08}.

For each observation and detector, we first built the source+background (extracted from the $10\arcsec$-radius region) and the background (extracted from the $3\arcmin \times 3\arcmin$ region)  light curves in the 2$-$10 keV energy range with 300 s time bins.
During this operation, we used the \texttt{epiclccorr} task to apply relative corrections to those light curves.
The relative corrections specify the good time intervals (GTI) of the event list according to the corresponding CCD and compute the livetime, i.e., select the time inside each CCD frame where the events were collected effectively (no FIFO reset/overflow, minimum ionizing particles, or read-out-time). 
Then, this task subtracts the background light curve (scaled to the same source extraction area) from the source+background light curve and scales up count rates and errors affected by the lost of exposure.
Finally, the background-subtracted light curves of the three detectors were summed to produce the EPIC light curves. 
Any missing count rate in a detector was inferred by the one observed by the other detectors using a scaling factor between them.
To do this, we calculated the scaling factor between the detectors during a time period where all three cameras were turned on.
The pn count rate is 1.31 times the sum of the MOS count rates.

\section{Timing analysis}
\label{timing}
\subsection{Bayesian blocks analysis}
\label{bb}
To identify the flaring and non-flaring levels under a certain probability using {\sl the unbinned} event arrival time, we used the Bayesian blocks analysis proposed by \citet{scargle98} and recently improved by \citet{scargle13b}.
The Bayesian blocks analysis of an event arrival time list from one of the EPIC cameras allowed us to segment the observing period with statistically different count rate levels and created a succession of constant count rate blocks.
The time defining two successive blocks is called a change point.
The count rate within each block is simply the number of events it contains divided by its effective exposure (livetime). 
The non-flaring and flaring levels are identified as the lowest and higher blocks, respectively.
The duration of the flares are determined as the time range of the Bayesian block corresponding to the elevated count rate.
This algorithm gives us the duration of the flaring and non-flaring levels with better accuracy than in a binned light curve since it uses the best temporal resolution available.

The number of change points is controlled by two input parameters: the false detection probability ($p_\mathrm{1}$) and the prior estimate of the number of change points, $ncp\_prior$.
We use $p_\mathrm{1} = \exp(-3.5)$ \citep{neilsen13,nowak12}, i.e., the probability that a found change point is a real change point is $1-\exp(-3.5)=96.8\%$ and the probability that a flare (at least two change points) is a real flare is $1-p_\mathrm{1}^2 = 99.9\%$.
We cannot use the geometric prior of \citet{scargle13b} since our data contain more events and our non-flaring level is lower than in the simulations used by \citet{scargle13b}, see Appendix \ref{appendix_a} for a detailed explanation.

We used the EXPOSU\#\# extension (\#\# corresponding to the CCD number where the source extraction region is located) of the event list to compute the detector live time from the nominal TIMEDEL of the corresponding instrument, i.e., the integration time without the time of the shift of a CCD line to the readout node, which is about 0.0687022 and 2.59 s for pn and MOS, respectively.
The Bayesian-block algorithm is used on the list of event of the source+background and of the background in which we selected the GTI (i.e., we reject the time where the camera did not observe).
It allows us to correct the light curve source+background from the flaring background following the recipe of \citet{scargle13}.
Indeed, thanks to Bayesian blocks, we know what the background count rate is and where the high-background levels are.
We can correct the source+background event list of any background contribution by applying a weight to each event, which is $w=CR_{\mathrm{src+bkg}}/(CR_{\mathrm{src+bkg}}+CR_{\mathrm{bkg}})$ with $CR_{\mathrm{src+bkg}}$ the count rate of the Bayesian blocks of the source+background event list and $CR_\mathrm{bkg}$ the count rate of the corresponding Bayesian block of the background that is surface-corrected \footnote{With this recipe we keep all the source+background events, sowe do not have to remove (arbitrarily) some individual events from the event list as proposed by \citet{stelzer07b} to subtract the background events.}.
Then, the Bayesian-blocks algorithm is applied a second time to this corrected source+background event lists.
This method is used on the three cameras separately.

\subsection{Smoothed light curve}
\begin{figure*}[!ht]
\centering
\includegraphics*[trim = 1cm 9cm 2cm 2cm, clip,width=16.9cm, angle=180]{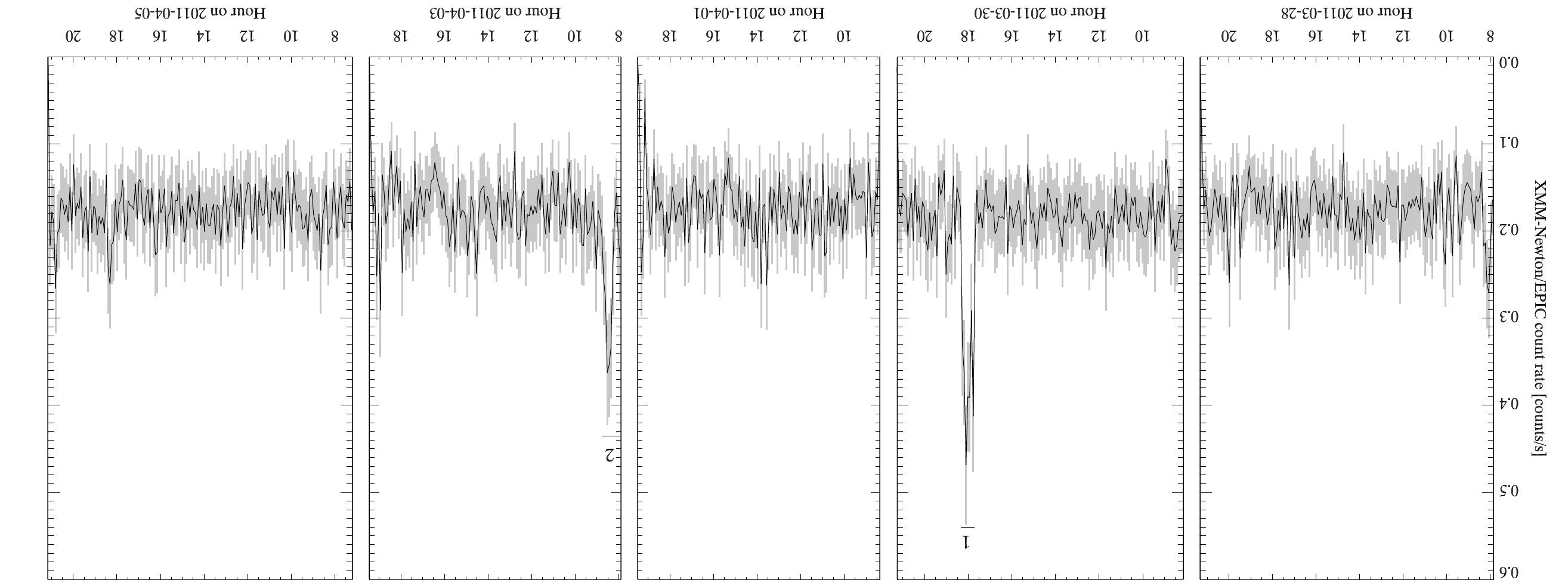}
\caption{XMM-Newton/EPIC (pn+MOS1+MOS2) light curves of \sgra{} in the 2–10 keV energy range obtained in Spring 2011. 
The time interval used to bin the light curve is 300 s.
The X-ray flares are labeled from 1 to 2.
The horizontal lines below these labels indicate the flare durations.
The non-flaring level of \sgra{} corresponds to only 10\% of the non-flaring level of these light curves \citep{porquet08}.}
\label{Fig1}
\end{figure*}

We compute a smoothed light curve by applying a density estimator \citep{silverman86,Feigelson12} on the {\sl unbinned} event arrival times using GTI to suppress camera switch-off.
The density estimator improves the characterization of light curve features, e.g., the amplitude and the time of a local maximum or minimum.
The density is computed using \texttt{quantreg} in R package, which convolves the event arrival times with a smoothing kernel. 
We modify \texttt{quantreg} to use the Epanechnikov kernel, which is defined as $K(x)=\frac{3}{4}\left(1-x^2\right)$ for $\mid x \mid \le 1$ and $K(x)=0$ for $\mid x \mid > 1$.
We chose the Epanechnikov kernel since it has "good performance" \citep{Feigelson12};
moreover, it is defined on a finite support, which allows us to control any boundary effects.
The density estimator can be expressed as
\begin{center}
\begin{equation}
 \hat{f}(t,h)=\frac{1}{N}\sum_{i=1}^N \frac{w(t)}{livetime} \times K\left(\frac{t-t_\mathrm{i}}{h}\right)
\label{eq1}
\end{equation}
\end{center}
\begin{figure}[!h]
\centering
\includegraphics*[trim = 10cm 0cm 5cm 0cm, clip,width=4.cm,angle=90]{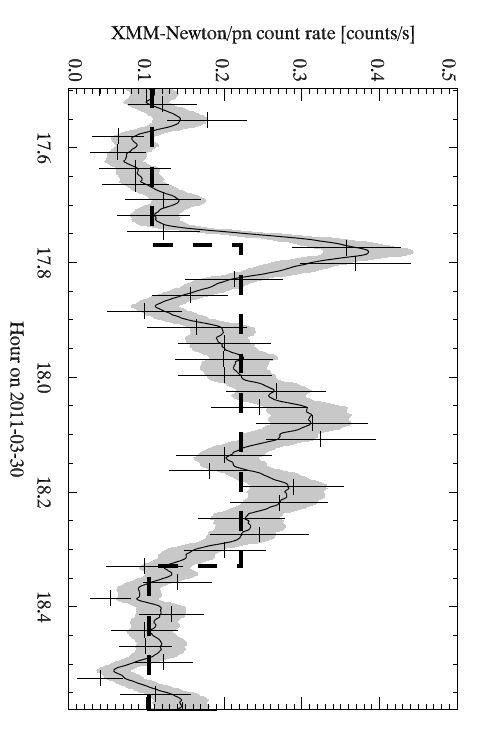}\\
\includegraphics*[trim = 15cm 0cm 0cm 0cm, clip,width=4.cm,angle=90]{24682_fig2.pdf}
\caption{XMM-Newton light curve of the 2011 March 30 flare from \sgra{} in the 2–10 keV energy range. 
\textit{Top panel:} The XMM-Newton/EPIC pn light curve binned on 100s.
The crosses are the data points of the light curve.
The horizontal dashed lines represent the non-flaring level found by the Bayesian-blocks algorithm.
The vertical dashed lines show the start and stop of the Bayesian block.
The solid line is the smoothed light curve.
The gray curves are the errors associated with the smoothed light curve.
\textit{Bottom panel:} The XMM-Newton/EPIC (pn+MOS1+MOS2) light curve binned on 100s.
The horizontal dashed line represents the non-flaring level calculated as the sum of the non-flaring level in each instrument found by the Bayesian blocks.
The vertical dashed lines represent the beginning and the end of the flare calculated by the Bayesian-blocks algorithm on pn camera.
The solid line is the smoothed light curve, which is the sum of the smoothed light curve for each instrument (calculated on the same time range).
The gray curves are the errors associated with the smoothed light curve.}
\label{Fig2}
\end{figure}
\begin{figure}[!h]
\centering
\includegraphics*[trim = 10cm 0cm 5cm 0cm, clip,width=4.cm,angle=90]{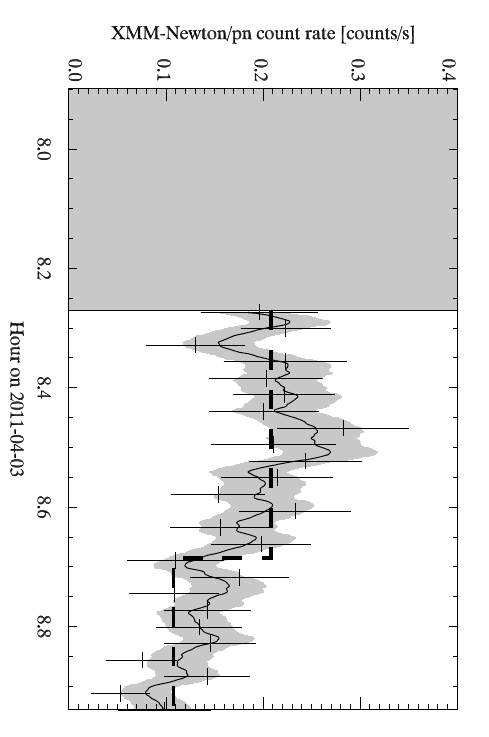} \\
\includegraphics*[trim = 15cm 0cm 0cm 0cm, clip,width=4.cm,angle=90]{24682_fig3.pdf}
\caption{XMM-Newton light curve of the 2011 April 3 flare from \sgra{} in the 2–10 keV energy range. 
\textit{Top panel:} The XMM-Newton/EPIC pn light curve binned on 100s.
The crosses are the data points of the light curve.
The horizontal dashed lines represent the non-flaring level found by the Bayesian-blocks algorithm.
The vertical dashed lines show the start and stop of the Bayesian block.
The solid line is the smoothed light curve.
The gray curve are the errors associated with the smoothed light curve.
The time period during which the camera did not observe is shown with a light gray box.
\textit{Bottom panel:}  The XMM-Newton/EPIC (pn+MOS1+MOS2) light curve binned on 100s.
The horizontal dashed line represents the non-flaring level calculated as the sum of the non-flaring level in each instrument found by the Bayesian blocks.
The vertical dashed lines represent the beginning and the end of the flare calculated by the Bayesian-blocks algorithm on pn camera.
The solid line is the smoothed light curve, which is the sum of the smoothed light curve for each instrument (calculated on the same time range).
The gray curve are the errors associated with the smoothed light curve.}
\label{Fig3}
\end{figure}
Then, we insert observing gaps using GTIs and combine the light curves of the three instruments.
with $h$ the width of the kernel window, $N$ the number of count in the event list, $t_\mathrm{i}$ the arrival time of the event $i$ and $t$ the time at which we compute the smoothed light curve, $w(t)$ the weight that corrects the density at the time $t$ from the flaring background thanks to the Bayesian-blocks algorithm (see above Sect.~\ref{bb}) and $livetime$ is the live time in the time interval [$t_\mathrm{i}-h/2$, $t_\mathrm{i}+h/2$].
The time $t$ is chose by the user.
Here, we take an even time grid with point interval of 5 s\,\footnote{The position of local extrema can be easily computed with required accuracy directly from the first derivative of Eq. \ref{eq1}.}.
We choose a constant window width of the kernel $h=100$ s.
The smoothed count rate ($CR$) is obtained from the density by $CR=N \, \hat{f}(t,h)$.
The error of the smoothed light curve is assumed to be Poissonnian ($=\sqrt{n}$ with $n$ the number of count in the kernel window).

\subsection{Results}
The EPIC (pn+MOS1+MOS2) background-subtracted light curves of \sgra{} in the 2$-$10 keV energy range, with a time bin interval of 300 s, are shown in Fig. \ref{Fig1}.
Our Bayesian-blocks analysis of the event list for individual detectors shows that during the first exposure, no flares were detected and the activity of the source region is constant. 
The first flare (\#1) was observed on 2011 March 30 and the last flare (\#2) on 2011 April 3.
The non-flaring level is determined as the count rate of the longest time interval of the Bayesian blocks of the non-flaring level, which allows very good accuracy on the count rate of the non-flaring level. 
On 2011 March 28, March 30, April 1, April 3, and April 5, the total non-flaring level was equal to $0.179 \pm 0.003$, $0,185 \pm 0.004$, $0.177 \pm 0.003$, $0.183 \pm 0.004\ \mathrm{count}\ \mathrm{s^{-1}}$, and $0.179 \pm 0.003\ \mathrm{count}\ \mathrm{s^{-1}}$, respectively.
It is consistent with the one previously observed with XMM-Newton (e.g., in 2007, \citealt{porquet08}).
This non-flaring emission is a combination of emission coming from the complex of stars IRS 13, the candidate pulsar wind nebula G359.950.04, and a diffuse component, which all contribute 90\% of this non-flaring level in the 2$-$10 keV energy range \citep{baganoff03,porquet08} and the emission from \sgra{} which contribute only 10\%.
Figures \ref{Fig2} and \ref{Fig3} focus on the flare light curves obtained with EPIC (pn+MOS1+MOS2) and EPIC pn with a bin time interval of 100s.
The comparison with the EPIC MOS1 and MOS2 light curves can be found in Appendix \ref{appendix_b}. 
We also show the Bayesian block corresponding to each camera with a dashed line.
Table \ref{table:2} gives the characteristics of these X-ray flares. 

\begin{table*}[!t]
\caption{Characteristics of the X-ray flares observed by XMM-Newton/EPIC in 2011.}
\centering
\scalebox{.9}{
\label{table:2}
\begin{tabular}{c c c c c c c c c c}
\hline
\hline
Flare & Day & Start Time\tablefootmark{a} & End Time\tablefootmark{a} & Duration & Total\tablefootmark{b} & Peak\tablefootmark{c} & $L\mathrm{^{unabs}_{2-10~keV}}$\tablefootmark{d} \\
(\#) & (yy-mm-dd) & (hh:mm:ss) & (hh:mm:ss)  & (s) & (cts) & ($\mathrm{count}\ \mathrm{s^{-1}}$) & ($10^{34}\ \mathrm{erg}\ \mathrm{s^{-1}}$) \\
\hline
 1 & 2011-03-30 & 17:46:20.69 & 18:19:40.86 & 2000.16 & $211 \pm 25$ & $0.284 \pm 0.013$ & $2.69 ^{+2.4}_{-0.7}$ \\
 2 & 2011-04-03 & $\le$08:16:35.65 & 08:41:02.04 & $\ge$ 1457.67 & $\ge 154 \pm 24$ & $0.165 \pm 0.012$ & $\ge 2.9 $ \\
\hline
\end{tabular}
}
\tablefoot{
\tablefoottext{a}
{Start and end times (TT) of the flare time interval defined by the Bayesian-blocks algorithm \citep{scargle13b} on the EPIC/pn data;}
\tablefoottext{b}
{Total EPIC/pn counts in the 2–10 keV energy band obtained in the smoothed light curve during the flare interval (determined by Bayesian blocks) after subtractingf the non-flaring level obtained with the Bayesian-blocks algorithm;}
\tablefoottext{c}
{EPIC pn count rate in the 2–10 keV energy band at the flare peak (smoothed light curves) after subtracting the non-flaring level;}
\tablefoottext{d}
{Unabsorbed 2–10 keV average luminosity of the flare computed from the total counts collected during the flare (i.e., the average count rate) and} assuming a distance of 8 kpc, see Sect. 4 for details.
}
\normalsize
\end{table*}

The first flare has two components: a short ($\sim 458$ s) and symmetrical subflare and a longer ($\sim 1542$ s) and fainter symmetrical subflare.
Between these two subflares, the smoothed light curve returns at 17.87h and during less than 100s to a lower level, which is consistent with the non-flaring state.
The first flare is seen in EPIC MOS1 camera with a shift of $\approx 75\ \mathrm{s}$ of its maximum at the first peak but the double subflare configuration is not seen in the EPIC MOS2 camera.
The amplitude of the flare in the smoothed light curve corresponds to 6.8 $\sigma$ (the standard deviation of the non-flaring level in the 300 s binned light curve) after subtracting the non-flaring level computed by the Bayesian-blocks algorithm.

The second flare is seen by the Bayesian-blocks algorithm in pn and EPIC MOS2 cameras but not in MOS1.
This can be explained by the detection limit of the algorithm and the lower sensitivity of the MOS cameras (see details in Appendix \ref{appendix_b}).
The time start of this flare is lower than or equal to the start of the observation. 
In EPIC MOS2 camera we can see an enhancement of the count-rate level after 8.25 h on April 3 but the Bayesian block algorithm detected also an enhancement at the beginning of the observation (before 8 h). 
Because of the time delay of observation start of EPIC pn camera, we caught with this camera only the end of this flare \footnote{A coordinated near-infrared observation was obtained with VLT/NACO on 2011 April 3 from 06:30 to 08:18 UT (ESO’s archive), which detected the rise of this flare (S. Gillessen, 2011, private communication), but lead to only 22 min of simultaneous observation with EPIC MOS before the flare peak that we observed in X-rays.}.
The amplitude of the flare subtracted from the non-flaring level corresponds to 5.9 $\sigma$.

We also computed the hardness ratio using the $2-4.4\ \mathrm{keV}$ and $4.4-10\ \mathrm{keV}$ energy bands for all observations, but we found no significant spectral change during the flare interval.
The peak count rates of the first and second flares are three and eight times less than that of the bright flare reported in \citet{porquet03}. 
The durations of theses flares are 1.4 and 1.8 times shorter than this bright flare. 

\section{Spectral analysis}
\label{spectrum}
We did a spectral analysis of the first flare.
The extraction region is the same as the one we used to construct the light curves, i.e., a circle of 10$\arcsec$ radius centered on the \sgra{} radio position.
The spectrum analysis was only done on the pn instrument since the flare in MOS1 and MOS2 has a number of counts that is too small to constrain the spectral properties. 
The X-ray photons were selected with \texttt{PATTERN$\leq 4$} and \texttt{FLAG==0}.
The time interval of the flare was constrained by the results of the Bayesian-blocks algorithm (see Table \ref{table:2}).
The background time interval is composed of two subintervals: the first one began at the start of the March 30 observation (see Table \ref{table:1}) and ended 300s before the beginning of the flare to avoid any bias.
The second one started 300s after the end of the flare and stopped at the end of the March 30 observation.

The spectrum, response matrices, and ancillary files were computed with the SAS task \texttt{especget}.
We used {\tt ISIS} version 1.6.2-27 \citep{houck00,houck02} to fit the spectrum with X-ray emission models.
The model that we used is an absorbed power law with dust scattering (\textit{dustscat}; \citealt{predehl95}).
NuSTAR observations of \sgra{} confirm that this model is still a good description of the flare spectra above 10 keV \citep{barriere14}.
We used \textit{TBnew} for the absorption model, the interstellar medium abundances developed by \citet{wilms00}, and the cross sections from \citet{verner96}.
These lower metal abundances and updated cross sections imply decreasing the column density \citep{nowak12} input to the \textit{dustscat} model \citep{predehl95}
by a factor of 1.5 times. It uses the $N_\mathrm{H}$ vs. $\tau_\mathrm{scatt}$ relation obtained with \textit{wabs} (\citealt{anders82}'s abundances, \citealt{morrison83}).

\begin{figure}
\centering
\includegraphics*[width=4.3cm,angle=-90]{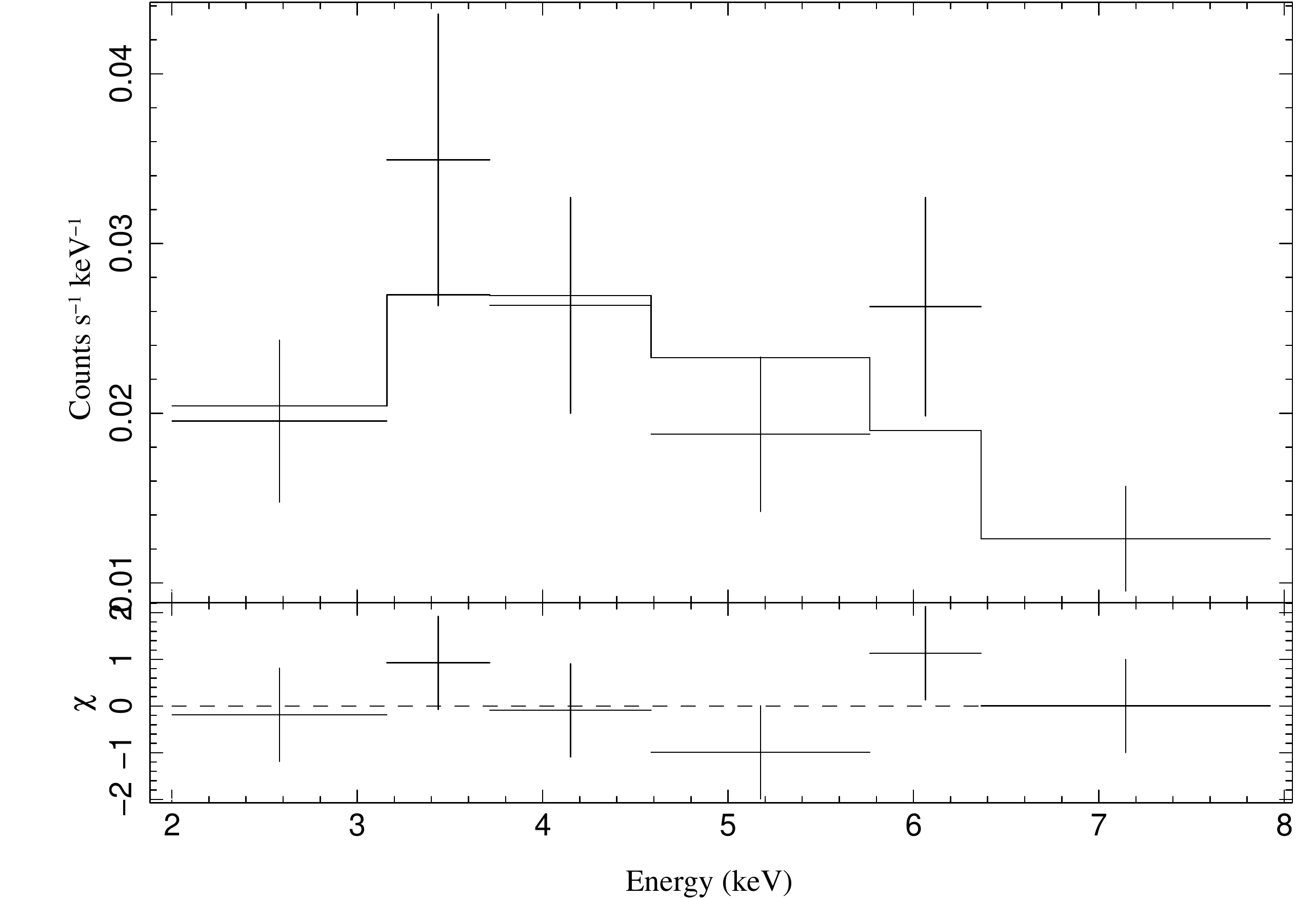}
\caption{XMM-Newton/EPIC pn spectrum of the 2011 March 30 flare.
The data are denoted by crosses.
The vertical bars is the 1$\sigma$ error in the count rate and horizontal bars show the spectral bin in energy. 
The events have been grouped with a minimum signal-to-noise ratio of 4.
{\it Top:} The result of the fit is shown by the continuous solid line.
{\it Bottom:} The $\chi^2$ residual in units of $\sigma$.}
\label{fig:10}
\includegraphics[width=4.3cm,angle=-90]{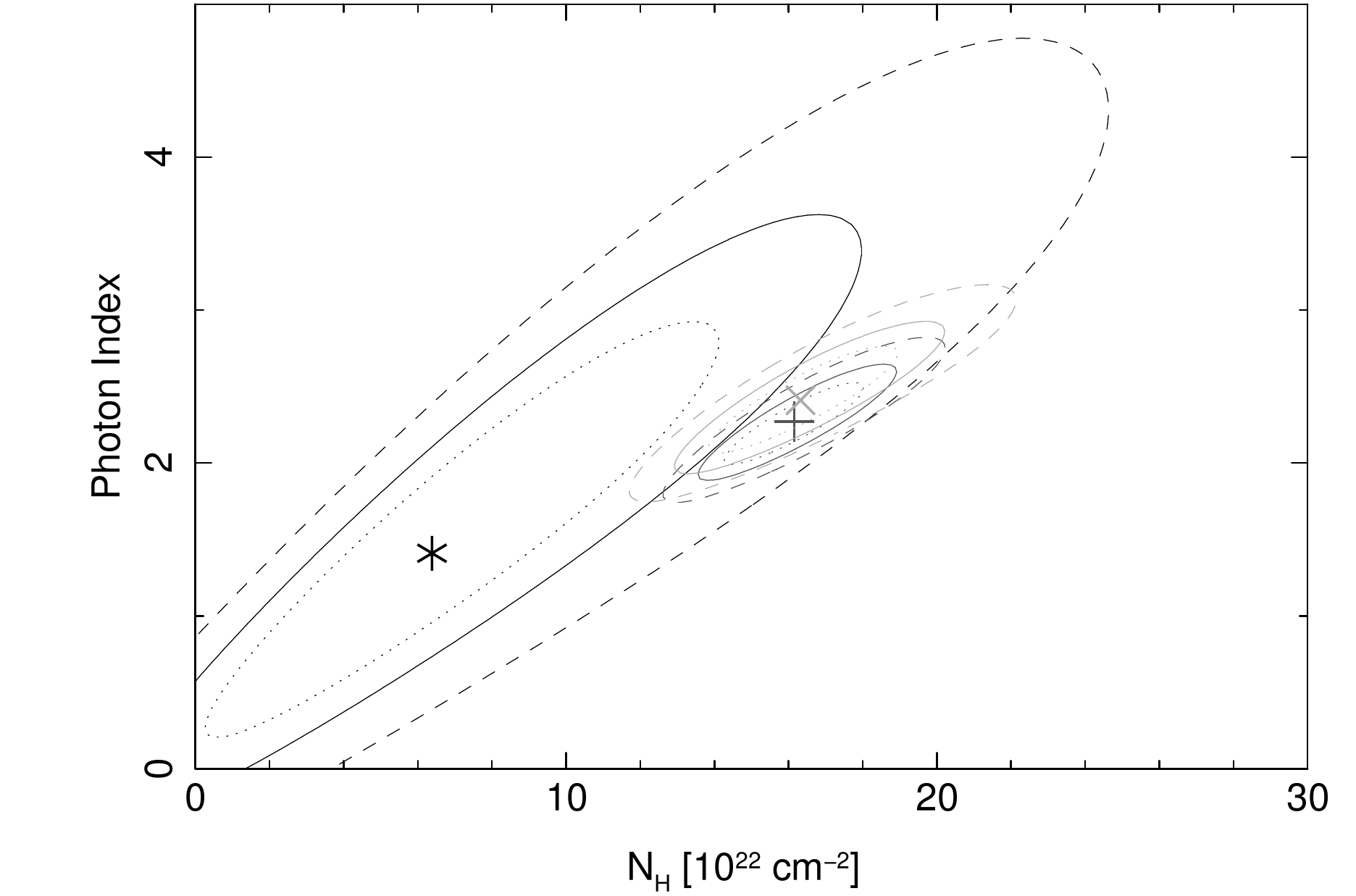}\\
\includegraphics[width=4.55cm,angle=-90]{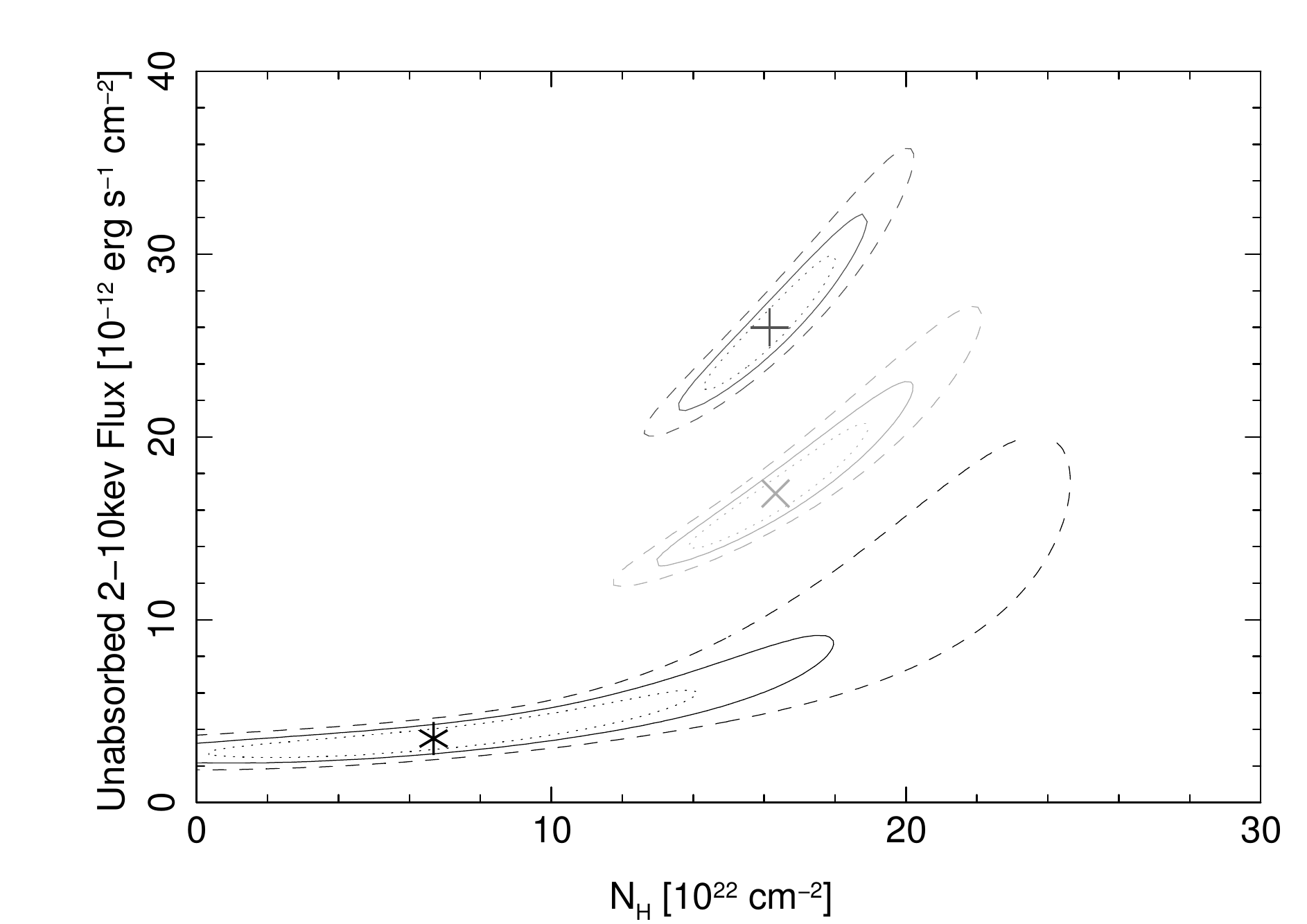}
\caption{Confidence contours for \sgra{} spectral parameters.
Contours are the confidence levels of 68\% (dotted line), 90\% (solid line), and 99\% (dashed line) for the two parameters in the graph.
The three sets of confidence contours represent the 2011 March 30 flare (black lines and asterisk), the 2007 April 4 flare (light  gray lines and X point), and the 2002 October 3 flare (dark gray lines and cross point).
}
\label{fig:9}
\end{figure}

The results of the fit using 90\% confidence level are hydrogen column density $(N_\mathrm{H})$ of $6.7^{+8.2}_{-6.7} \times 10^{22}\ \mathrm{cm^{-2}}$, photon index $(\Gamma)$ of $1.5^{+1.5}_{-1.3}$, absorbed flux between 2 and 8 keV $(F\mathrm{^{abs}_{2-8~keV}})$ of $2.5 \times 10^{-12}\ \mathrm{erg\ s^{-1}\ cm^{-2}}$ and unabsorbed flux between 2 and 10 keV $(F\mathrm{^{unabs}_{2-10~keV}})$ of $3.5^{+3.1}_{-1.0} \times 10^{-12}\ \mathrm{erg\ s^{-1}\ cm^{-2}}$.
The extracted spectrum and best fit are shown in Fig.~\ref{fig:10}.

We can compare the spectral parameters of this flare with those of the two brightest flares detected with XMM-Newton, which have the better constrained spectral parameters thanks to the high throughput and no pileup.
The very bright flare of 2002 October 3 was fitted using the same modeling with $\Gamma = 2.3 \pm 0.3$, $N_\mathrm{H} = 16.1^{+1.9}_{-2.2} \times 10^{22}\ \mathrm{cm^{-2}}$ , and $F\mathrm{^{unabs}_{2-10~keV}}=26.0^{+4.6}_{-3.5} \times 10^{-12}\ \mathrm{erg\ s^{-1}\ cm^{-2}}$ \citep{porquet03,nowak12}.
The bright flare of 2007 April 4 was also fitted using the same modeling with $\Gamma = 2.4 ^{+0.4}_{-0.3}$, $N_\mathrm{H} = 16.3^{+3.0}_{-2.6} \times 10^{22}\ \mathrm{cm^{-2}}$ and $F\mathrm{^{unabs}_{2-10~keV}}=16.8^{+4.6}_{-3.0} \times 10^{-12}\ \mathrm{erg\ s^{-1}\ cm^{-2}}$ \citep{porquet08,nowak12}.
In Fig.~\ref{fig:9}, the confidence contours of these two bright flares show that these $N_\mathrm{H}$ and $\Gamma$ parameters are well constrained.
However, those of the 2011 March 30 flare are not, since there are few events collected from this flare, which implies that the number of spectral bins with a minimum signal-to-noise ratio of 4 is small.
The photon index and hydrogen column of the flare of 2011 March 30  agree with those of the flare of 2007 April 4 and 2002 October 3 within the confidence levels for two parameters of 90\% and 99\%, respectively.

The second flare on 2011 April 3 does not contain enough counts to constrain the spectral parameters.
Its unabsorbed luminosity given in Table \ref{table:2} is calculated with {\tt ISIS} by fixing the photon index $\Gamma$ to 2 and column density to $N_\mathrm{H} = 14.3 \times 10^{22}\ \mathrm{cm^{-2}}$, i.e., to the spectral values of the 2002 October 3 flare \citep{porquet03,nowak12}.
Thus, the only free parameter is  the unabsorbed flux, which is $F\mathrm{^{unabs}_{2-10~keV}}=3.91 \times 10^{-12}\ \mathrm{erg\ s^{-1}\ cm^{-2}}$.
The unabsorbed luminosity between 2 to 10 keV is $2.7 ^{+2.4}_{-0.8}\times 10^{34} \ \mathrm{erg\ s^{-1}}$ for a 8 kpc distance.

\section{Discussions}
\label{discussion}
\subsection{The 2011 March 30 flare vs. the 2012 \textit{Chandra XVP} campaign flares}
\begin{figure}[!h]
\centering
\includegraphics[trim = 0cm 0cm 0.63cm 0cm, clip,width=6cm,angle=90]{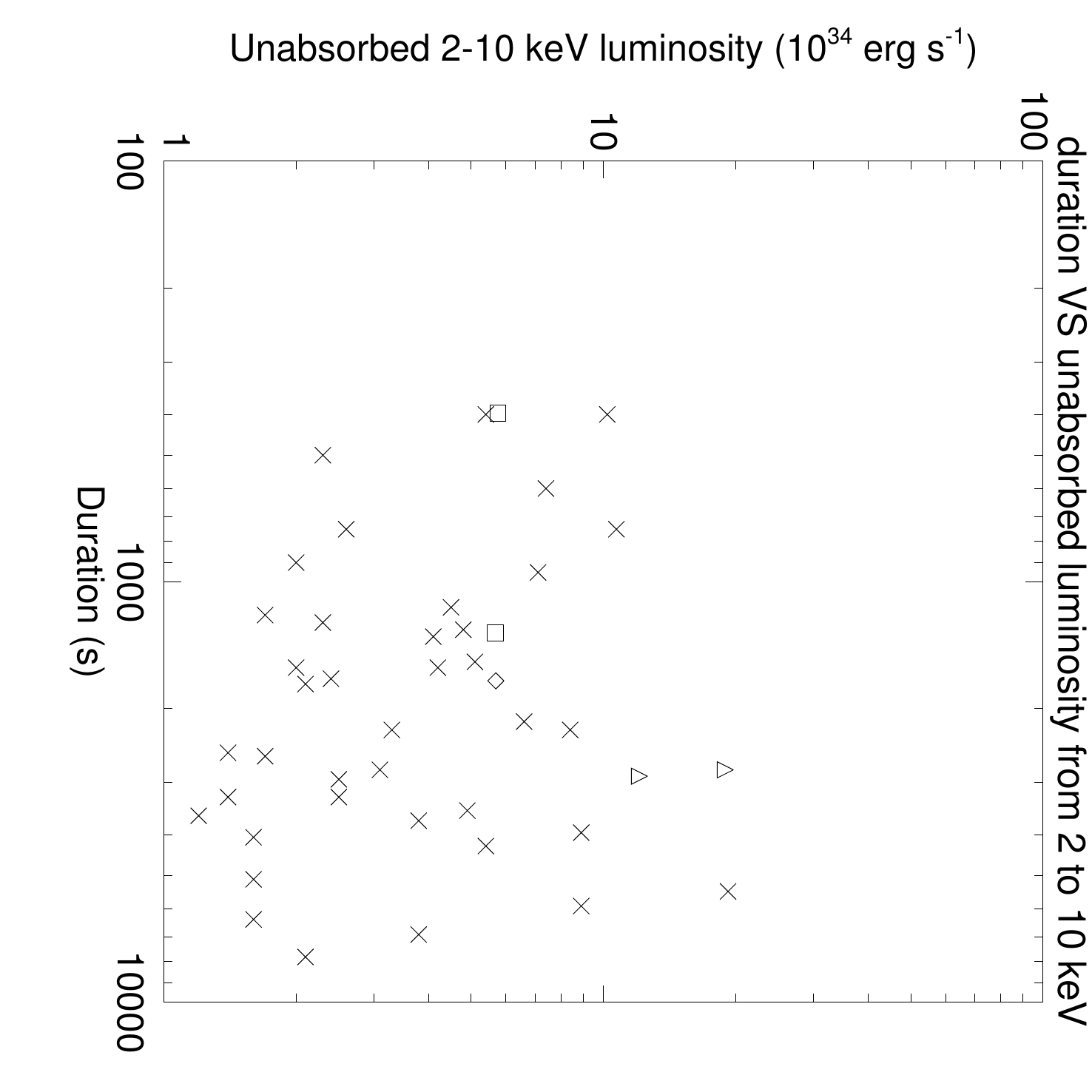}\\
\includegraphics[trim = 0cm 0cm 0.63cm 0cm, clip,width=6cm,angle=90]{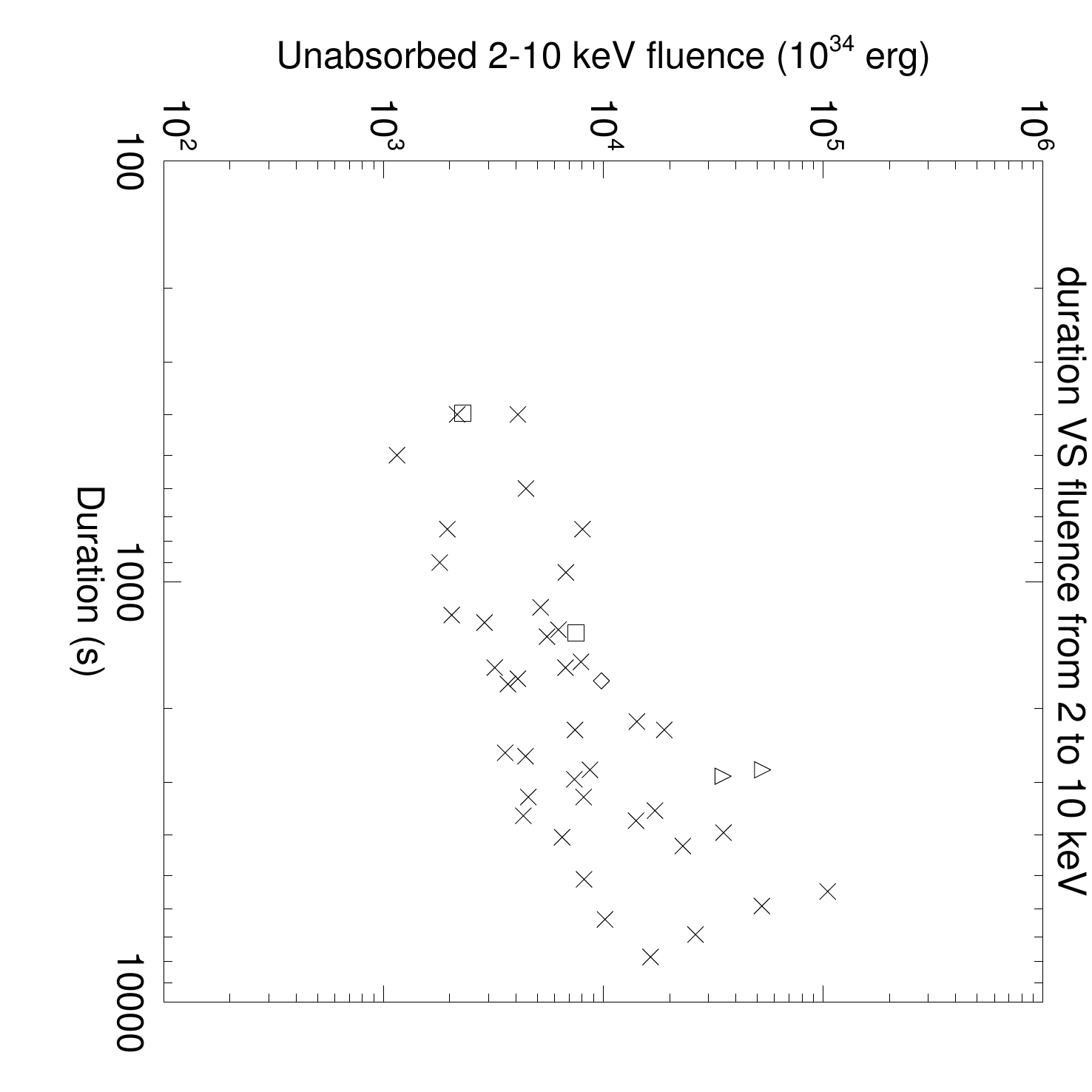}\\
\includegraphics[trim = 0cm 0cm 0.63cm 0cm, clip,width=6cm,angle=90]{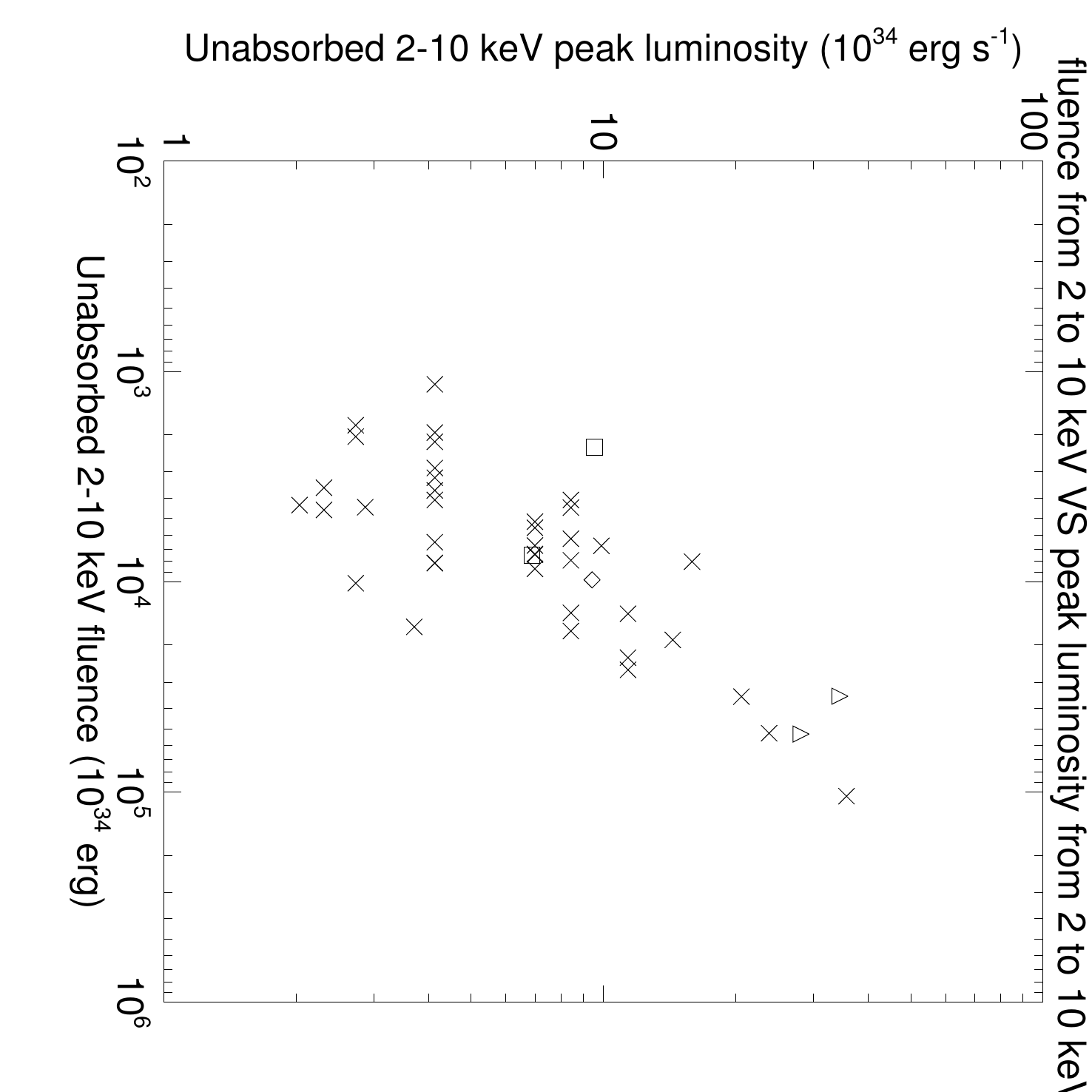}
\caption{2011 March 30 flare vs. the 2012 \textit{Chandra XVP} campaign flares.
The X-ray flares from the \textit{Chandra XVP} campaign \citep{neilsen13} are shown by crosses, the two brightest flares seen be XMM-Newton are triangles, the 2011 March 30 flare is represented by a diamond, and the two subflares are shown with squares.
}
\label{fig:11}
\end{figure}

We compared the spectral properties of the 2011 March 30 flare with the ones reported by \citet{neilsen13} from the 2012 \textit{Chandra XVP} campaign.
In this paper, the spectral properties of all Chandra flares have been derived by assuming the spectral parameters of the brightest flares obtained by \citet{nowak12}: $\Gamma=2$ and $N_\mathrm{H} = 14.3 \times 10^{22}\ \mathrm{cm^{-2}}$.
We use two physical quantities given in the Table 1 of \citet{neilsen13}: the unabsorbed 2$-$10 keV luminosity and the duration of the flare.
We also derived two other physical quantities that are independent of the instrumental characteristics: the unabsorbed 2$-$10 keV fluence in erg (the product of the unabsorbed 2$-$10 keV luminosity with the duration) and the unabsorbed 2$-$10 keV peak luminosity.
To compute the peak luminosity of the Chandra flares, we first derived the mean count rate in each flare as the pileup corrected fluence in counts (see Eq. 1 \citealt{neilsen13}) divided by the flare duration, and then we computed the linear relation between the unabsorbed 2$-$10 keV luminosity and the mean count rate (higher the mean count rate, higher the luminosity).
We obtained $L\mathrm{^{unabs}_{2-10~keV}}/\mathrm{10^{34}\ erg\ s^{-1}}=-0.031+136.7 \, (CR/\mathrm{count\ s^{-1}})$ with a correlation parameter $r$ of $0.9997$.
Then, we applied this relation to the peak count rate given in Table 1 of \citet{neilsen13} to obtain the peak luminosity for each flare.
The relations between these four physical quantities are shown in Fig \ref{fig:11}.

Since we used quantities that are independent of the instrument, we can compare flares observed with Chandra and XMM-Newton.
First, we place the two brightest flare seen by XMM-Newton in the three diagrams.
The unabsorbed 2$-$10 keV fluence, duration, and unabsorbed 2$-$10 keV luminosity are reported in \citet{nowak12}.
The unabsorbed 2$-$10 keV peak luminosity are computed as the ratio between the peak count rate and the mean count rate multiplied by the unabsorbed 2$-$10 keV luminosity in the flares \citep{porquet08}.
We can see that these flares have high luminosities and fluences. 
They are thus in the upper righthand corner of the diagrams representing the unabsorbed 2$-$10 keV peak luminosity and the unabsorbed 2$-$10 keV fluence.

\begin{table*}[!ht]
\caption{Characteristics of the 2011 March 3 flare and its two subflares assuming $\Gamma=2$ and $N_\mathrm{H} = 14.3 \times 10^{22}\ \mathrm{cm^{-2}}$.}
\centering
\scalebox{.78}{
\label{table:3}
\begin{tabular}{c c c c c c c}
\hline
\hline
Flare & Duration & Live time & Mean net count rate & $L\mathrm{^{unabs}_{2-10~keV}}$ & Peak count rate & $L\mathrm{^{unabs}_{2-10~keV}}$(peak)\\
(\#) & (s) & (s) & ($\mathrm{count\ s^{-1}}$) & ($10^{34}\ \mathrm{erg}\ \mathrm{s^{-1}}$) & ($\mathrm{count\ s^{-1}}$) & ($10^{34}\ \mathrm{erg}\ \mathrm{s^{-1}}$) \\
\hline
 1 & 2000 & 1750 & 0.16 & 5.7 & 0.28 & 9.5 \\
 1.1 & 458 & 416 & 0.16 & 5.8 & 0.28 & 9.4 \\
 1.2 & 1542 & 1324 & 0.16 & 5.7 & 0.17 & 6.8\\
\hline
\end{tabular}
}
\normalsize
\end{table*}

We also represent our first flare as a single flare (diamond) and as two distinct subflares (squares) defined as follows.
The first subflare starts at the beginning of the Bayesian blocks corresponding to the 2011 March 30 flare and stops at the time corresponding to the minimum of the smoothed light curve between the two subflares.
The beginning of the second subflare is the end of the first one, and its end corresponds to the end of the Bayesian block.
The live time of the single flare (Table \ref{table:3}) is shorter than the flare duration reported in Table \ref{table:2}.
The mean rate of the flare is given by the Bayesian-blocks algorithm.
The mean rate in each subflare is the number of counts in each subflare divided by their live time.
To be consistent with \citet{neilsen13} for direct comparison purpose, the unabsorbed 2$-$10 keV luminosity is computed with the same spectral parameter ($\Gamma=2$ and $N_\mathrm{H} = 14.3 \times 10^{22}\ \mathrm{cm^{-2}}$), which implies that the luminosity of the flare is slightly different than those computed with our present best fit spectral analysis (see Sect.~\ref{spectrum}).
The derived quantities for the first flare (\#1) and the first (\#1.1) and second (\#1.2) subflares are reported in Table \ref{table:3}.

In Fig. \ref{fig:11}, we can see that the unabsorbed 2$-$10 keV luminosity of the total flare and the two subflares are nearly the same since they have more or less the same mean count rate, but the fluence of the first sub-flare is small compared to the second subflare owing to the shorter duration.
Thus, the first subflare lies within the shortest and less energetic flares detected by Chandra, but the apparent lower detection limit of 400 s in the flare duration is probably due to the method used by \citet{nowak12} to identify flares in Chandra light curves.
In fact, they use a Gaussian fit on the light curve binned with 300 s, which implies that they might missed flares whose duration is below 300 s.

We can see that, in all the diagrams, if we assume a single flare, it lies in the mean of the flares seen by Chandra and can then be considered as a genuine medium luminosity flare.
Furthermore, if we consider the minimum waiting time between flares in the 2012 \textit{Chandra XVP} campaign shown in Fig. 1 of \citet{neilsen13}, we can see that the nearest flares are separated by $\sim 3500$ s.
This waiting time can be considered as a lower limit for observing two distinct flares.
The two subflare peaks of our first flare are separated with only $1000$ s, which favors a single flare.

\subsection{Gravitational lensing of a hotspot-like structure}
We modeled the light curve of the 2011 March 30 flare with a single mechanism in order to explain the two subflares.
Indeed, the very short ($\sim458$ s) first subflare and the second much longer ($\sim1542$~s) one peaking $\sim 1000$~s later but with lower amplitude can be the signature of a gravitational lensing of a hotspot-like structure.
We used a hotspot model and a ray-tracing code to compute the observed intensity \citep{karas92,schnittman03,broderick05,hamaus09,dexter09}.

\subsubsection{The hotspot model}
\label{hotspot}
We call a hotspot a spherical, optically thin structure, orbiting around the black hole with Keplerian angular velocity.
The sphere is initially assumed to be in solid rotation around the black hole. No shearing or expansion of the sphere is taken into account.
Such a hotspot is thus only defined by its radius $R$ and its orbital radius $r$ in gravitational radius unit ($r_{\mathrm{g}} \equiv 0.5 R_{\mathrm{S}}$). 
The black hole inclination $i$ is assumed to be close to an edge-on view, i.e., $i \approx 90^{\circ}$. 
Its actual value is a parameter of the model. 
The emitted spectrum of the hotspot is assumed to follow a power law, $I_{\nu}^{\mathrm{em}} \propto \nu^{\alpha}$, where $\alpha$ is a constant number, related to the photon index $\Gamma$ through $\Gamma = 2 - \alpha$. 
It is then straightforward to show that the observed intensity integrated over a range of frequency $\Delta \nu_{\mathrm{obs}}$ is
\begin{equation}
 \mathcal{I}^{\mathrm{obs}} = \int_{\Delta \nu_{\mathrm{obs}}} I_\nu^{\mathrm{obs}} \dd \nu_{\mathrm{obs}}  \propto g^{4-\alpha} 
\end{equation}
where $g \equiv \nu_{\mathrm{obs}} / \nu_{\mathrm{em}}$ is the redshift factor.

Maps of $\mathcal{I}^{\mathrm{obs}}$ were computed by using the open-source ray-tracing code \texttt{GYOTO}\footnote{This code can be freely downloaded at the URL \href{http://gyoto.obspm.fr}{gyoto.obspm.fr}} \citep{vincent11}.
We computed maps of $300 \times 300$ pixels over one orbital period with a time step of about $\delta t \approx 10$~s, which is close to the time sampling of the smoothed light curve. 
The light curve is obtained by summing each of these maps over all pixels, which boils down to integrating over all solid angles, i.e., to computing a flux.

The light curve of a hotspot seen edge-on shows a typical double-bump feature (see Fig.~\ref{fig:raddep}).
The primary maximum ($t=0$ s) is due to the gravitational lensing of the light emitted by the hotspot when it is on the opposite side of the black hole with respect to the observer. 
The secondary maximum ($t \approx 1000$ s) is due to the relativistic beaming effect: light emitted when the source is moving toward the observer is boosted.

\subsubsection{Constraining the orbital radius of the hotspot}
\begin{figure}
\centering
\includegraphics[width=4.3cm]{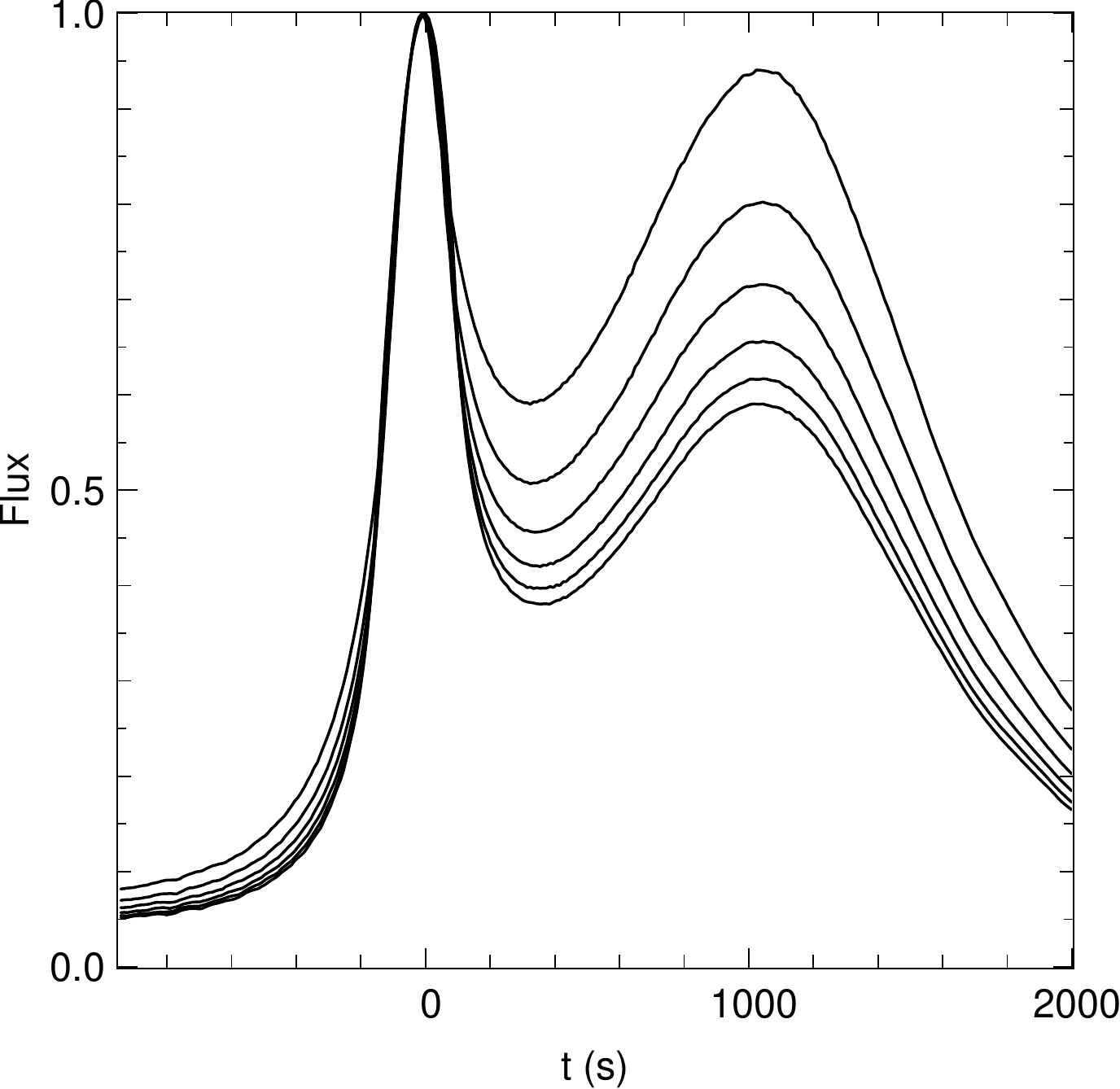}
\includegraphics[width=4.3cm]{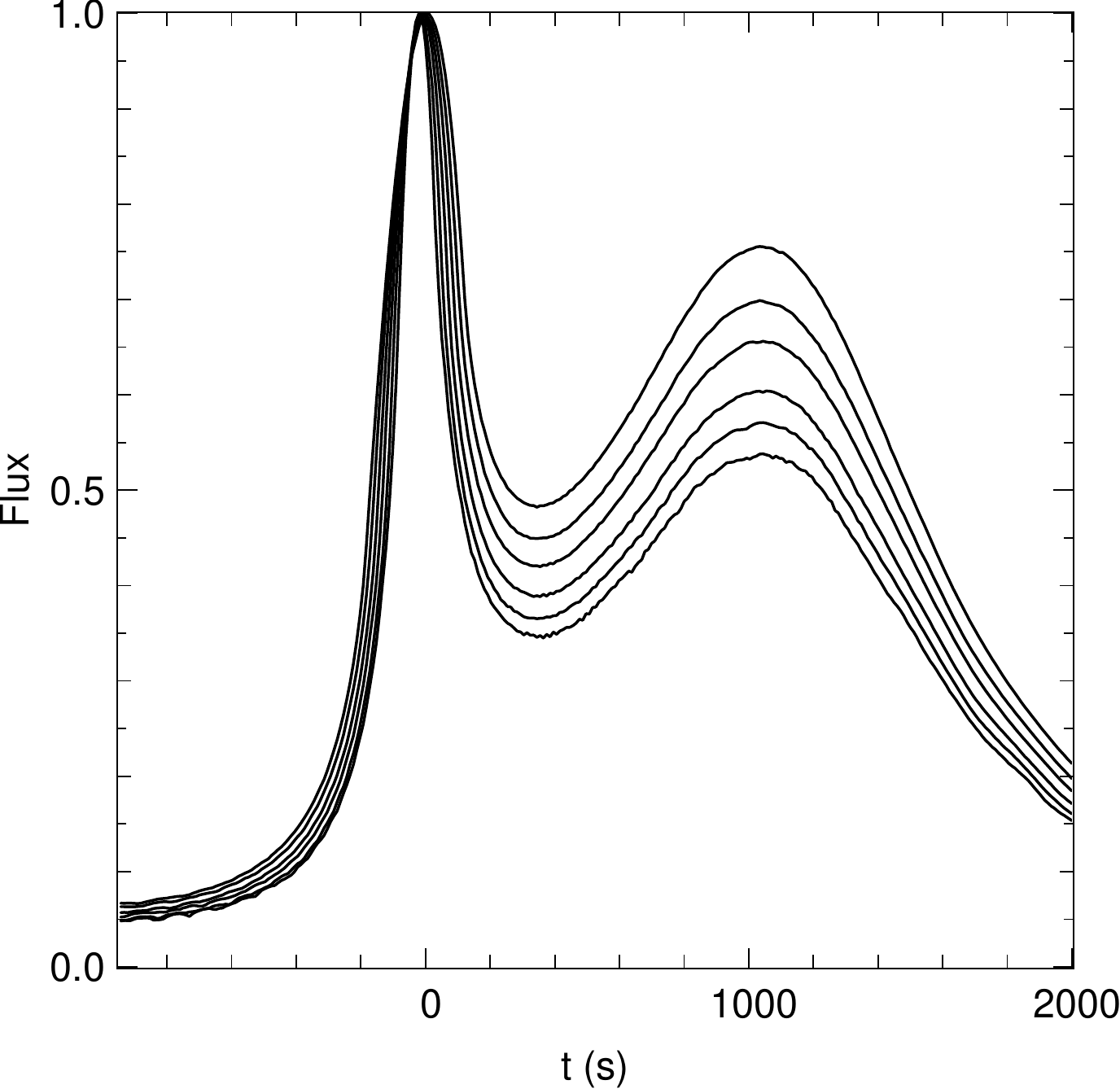}
\caption{Normalized hotspot light curves obtained for an orbital radius equal to the value of the best fit.
The flux is in arbitrary units.
{\it Top:} hotspot radius equal to the best fit value. Inclination $i$ varies over all grid values. (The range of these parameters are defined in Sect~\ref{r_gird}.)
The closer the inclination to $90^{\circ}$, the higher the ratio between the two local maxima.
{\it Bottom:} inclination fixed at its best fit value and $R$ varies over all grid values.
The smaller $R$, the bigger the ratio between the two subflare peaks.
The time interval between the two local maxima is the same for all curves in these two figures.}
\label{fig:raddep}
\end{figure}

The orbital radius is easy to constrain because it is directly linked to the time interval between the two local maxima of the light curve, as illustrated in Fig.~\ref{fig:raddep}. 
The variation of this time interval as a function of the orbital radius $r$ evolved like this: $\delta t \approx 860,\ 960,\ 1040,\ 1130,\ \mathrm{and}\ 1230$s for $r/r_\mathrm{g}=10.5,\ 11,\ 11.5,\ 12,\ \mathrm{and}\ 12.5$.
It is clear that if the hotspot model is correct, then $r/r_\mathrm{g}\approx$11$-$12.

It is not obvious to constrain the remaining parameters ($R$ and $i$) by a quick comparison to the observed data. 
Both of them have a strong impact on the flux ratio between the two local maxima, as well as on the flux ratio between the primary maximum and the local minimum between the two bumps.

\subsubsection{Fitting the parameters of the hotspot model}
\label{r_gird}
Our hotspot model is defined by five parameters: the orbital radius $r$, the hotspot radius $R$, the black hole inclination $i$, the temporal additive shift $dt$, and the flux multiplicative scaling $df$.
The two last parameters are defined according to the following.
The smoothed light curve defines the zero of time: it is by definition the time of its primary maximum.
Then, each theoretical light curve is first shifted so that its zero of time corresponds to its own primary maximum. 
The parameter $dt$ allows the fitting of any time shift between the  theoretical and the observed light curve.  
Each theoretical light curve is also scaled vertically. 
Each of them is first divided by the maximum of all fluxes computed by \texttt{GYOTO} (then all \texttt{GYOTO} fluxes are between $0$ and $1$).
Each theoretical light curve is then again multiplied by the maximum value of the smoothed light curve, from which the non-flaring ground level was subtracted. (Then all \texttt{GYOTO} flux values are between $0$ and $M$, the maximum of the smoothed light curve, in observed unit.) 
The multiplicative $df$ fitting parameter is applied to these rescaled theoretical light curves.

The spin parameter has a low impact on the light curve, thus it is fixed to $a=0.99$ (high spins lead to slightly smaller $\chi^2$ in the fit) and not fitted. 
The photon index is fixed to $\Gamma=2$ \citep{porquet03,porquet08,nowak12,barriere14}.
Here we are interested in determining whether the hotspot model is viable or not, not in fitting in detail all the parameters.

The fitting is performed by determining the minimum of the following $\chi^2$ on a grid of parameters
\fontsize{7.5}{7.5}\selectfont
\begin{equation}
 \chi^2(r,R,i;dt,df) = \sum_{t_{\mathrm{obs}}} \left(\frac{df\times f_{\mathrm{Gyoto}}(r,R,i;dt;t_{\mathrm{obs}})+f_{\mathrm{non-flaring}}-f_{\mathrm{smooth}}(t_{\mathrm{obs}})}{\sigma_{\mathrm{smooth}}}\right)^2
\end{equation}
\normalsize
where $f_{\mathrm{Gyoto}}$ is the theoretical light curve, $f_{\mathrm{non-flaring}}$ is the non-flaring level of the observed data (determined from the Bayesian-blocks analysis), $f_{\mathrm{smooth}}$  the pn smoothed light curve, $\sigma_{\mathrm{smooth}}$  the error on the smoothed flux, and the sum is performed over a subset of the range of observed times taken into account in the smoothing procedure, with a time step of about $10$~s.
We use conservatively only the pn smoothed light curve since pn is the most sensitive instrument.
The grid that we use for the three physical parameters is $r\in[10.5,11,11.5,12,12.5]$, $R\in[1.2,1.4,1.6,1.8,2,2.2],$ and $i\in[81.93,83.08,84.22,85.37,86.52,87.66]$ where radii are in units of $GM/c^2$, the inclination is in degrees, with $i=90^{\circ}$ being an exact edge-on view (i.e., maximum lensing effect).
For each set of parameter values, the theoretical light curve corresponding to $(r,R,i)$ is read. 
It is rescaled as described above. Then the parameters $dt$ and $df$ are fitted using the \texttt{lmfit} routine of the \texttt{Yorick} software. 
The set of parameters that gives the smallest $\chi^2$ following this procedure is the best-fitting set. 
For the fitting,  the theoretical light curve is fitted to the smoothed data, interpolating linearly to determine the theoretical value at the smoothed times. 

\begin{figure}
\centering
\includegraphics[trim = 5.9cm 0cm 5.5cm 0cm, clip, angle=90]{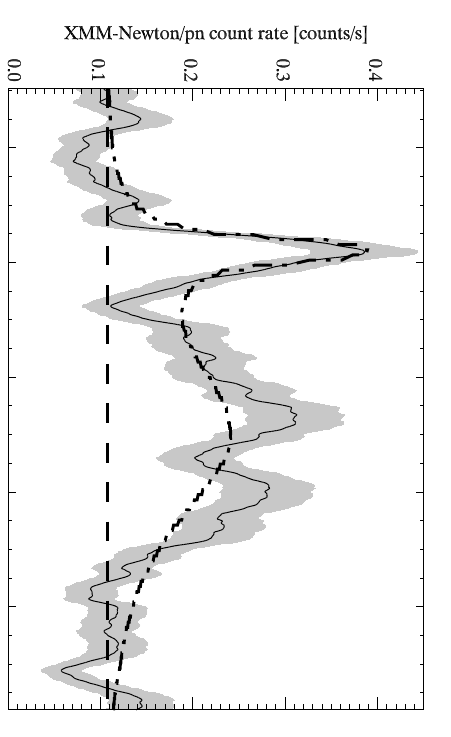}\\
\includegraphics[trim = 2.5cm 0cm 11cm 0cm, clip, angle=90]{24682_fig8.pdf}
\caption{Modeling of the 2011 March 30 flare pn light curve with a rotating hotspot.
Best-fitting theoretical light curve (dot-dashed line) plotted over the smoothed light curve (solid line, with $1 \sigma$ error in gray). 
The non-flaring level is given by the horizontal dashed line. 
The  vertical axis is in observed units, horizontal axis is in seconds. 
The lower panel gives the residual in units of $\sigma$.}
\label{fig:best}
\end{figure}

Figure~\ref{fig:best} shows the best fit that is found for the following values of the parameters: $r=12r_{\mathrm{g}}, R=1.4r_{\mathrm{g}}, i=86.5^{\circ}, dt= 11.1\pm4.0 \; \mathrm{s} , df= 1.40 \pm 0.02$.
The $1 \sigma$ error on the two last parameters being computed by the \texttt{lmfit} routine.
The final reduced $\chi^2$ is $0.85$.

\subsubsection{Viability of the hotspot model}
The best fit illustrated in Fig.~\ref{fig:best} clearly shows that one part of the smoothed data is not well fit by the hotspot model: the local minimum of the light curve, in between the two bumps at around 17h52m34s. 
At this point, the observed data reach the non-flaring level while the model stays much higher, around $2.5~\sigma$ distant.
The remaining data is explained well by the model at the $1~\sigma$ level. However, this $2.5\,\sigma$ inadequacy of the model at the local minimum is sufficient to reject the model. 
Indeed, a hotspot-like model will always produce a local minimum at a
higher level than the non-flaring level.
Indeed, this part of the light curve is associated to the part of the trajectory where the hotspot is moving from behind the black hole to the approaching side of the orbit. 
At this position, the relativistic beaming effect will always be significantly greater than at the receding side of the orbit, which corresponds to the minimum flux level. 

To be quantitative, we compare the flux ratio between the lensing maximum and the non-flaring level and the lensing maximum and the local minimum flux level (in between the two bumps), for all light curves computed in our grid.
The first ratio is always greater than ten, while the second ratio varies between $1.5$ and $4$.
In conclusion, no set of parameters can give the same ratio for these two quantities.

\subsubsection{Refining the hotspot model}
\label{refine_hotspot}
\begin{figure}
\centering
\includegraphics[width=6cm]{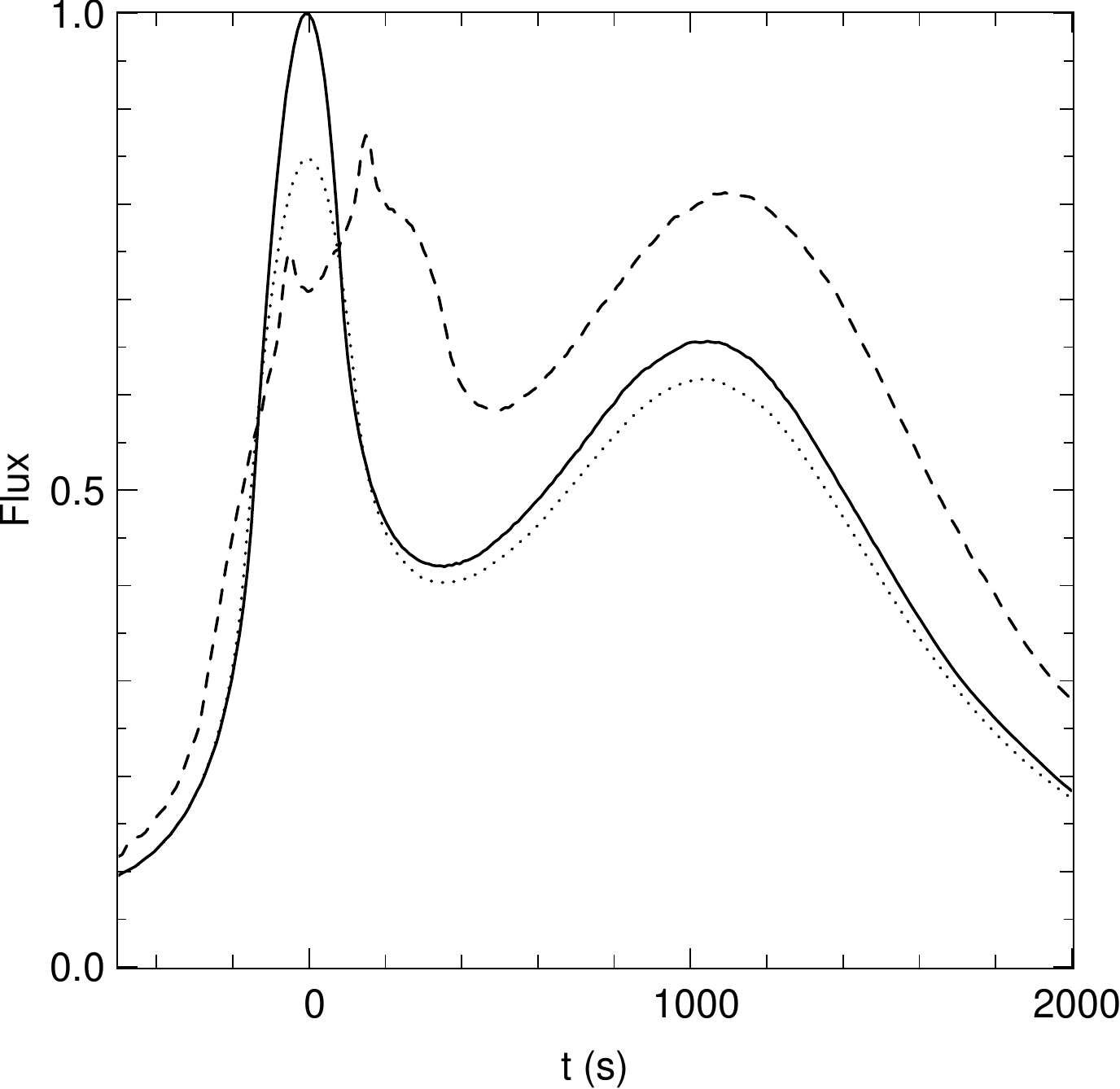}
\caption{Comparison of the best-fit light curve (solid line) with the light curve of an elongated (dashed line) and swelling (dotted line) hotspot defined by the same parameters (see text for details).
The flux is in arbitrary units.}
\label{fig:compare}
\end{figure}

One may wonder whether, by adding more physics to the hotspot model, this local minimum problem could be solved. 
To investigate this, we considered the two most natural ways of making our model more sophisticated: considering an elongated hotspot due to the shearing of the sphere by the differential Keplerian rotation and allowing the hotspot to vary in radius ($R$) along its trajectory. 
To model an elongated hotspot, we first computed the effect of elongation over a hotspot of initial radius $R=1.8$ over the time elapsed between the triggering of the hotspot and the local minimum. 
The precise triggering time of the hotspot is not constrained, thus we assume the hotspot is created at the time that corresponds to the minimum of the theoretical light curve (thus at about $-1500$~s when $t=0$ is set at the primary maximum). 
Under this assumption, the elapsed time between the creation of the hotspot and the local minimum is of $\Delta T \approx1800~$s for the best-fitting values of parameters. 
This is equivalent to one third of the period.
It is now straightforward to compute the difference of angular distance $\Delta \theta$ covered by the most distant (in terms of radial coordinate $r$) and least distant parts of a sphere with radius $R=1.8$ whose center is at a radius $r=11.5$ from the central black hole. 
Explicitly,
\begin{equation}
 \Delta \theta = \frac{2\pi}{3}\,r^{3/2}\left(r_1^{-3/2}-r_2^{-3/2}\right)
,\end{equation}
where $r_1$ is the shortest distance to the black hole and $r_2$ the largest (the spin is neglected here).
For the best-fitting hotspot, $\Delta \theta \approx \pi/3$. 
This elongation has approximately multiplied the angular extension of the initially spherical best-fit hotspot by three. 
Thus, we model the elongated hotspot in a simple way by considering three spherical best-fit hotspots tangent one to the next, orbiting the same orbit. 
The intensity emitted by each of the spheres is divided by three with respect to the standard single hotspot case in order to allow a simple comparison. 
Figure~\ref{fig:compare} shows the light curve associated to this elongated hotspot. 
Here, the hotspot is always elongated and does not change shape as a function of time. 

To determine the effect of volume changing on the light curve, we have modeled a swelling, single hotspot.
The swelling hotspot is modeled by requiring that the initial radius of the hotspot is the best-fitting value, $R_0=1.8$ and that it will increase linearly with time until it reaches $2\,R_0$ at the time corresponding roughly to the local minimum observation. 
The emitted intensity is inversely proportional to the sphere volume.
Figure~\ref{fig:compare} shows the light curve associated to this swelling hotspot.

Both toy models show that changing the shape of the hotspot will not solve the central problem of the model: the local minimum is always significantly higher than the non-flaring level.
We cannot formally exclude that a more sophisticated model, such as a non-constant density hotspot, or a trajectory not confined within the equatorial plane is able to fit our data. 
However, it is important to note that our hotspot model is ruled out precisely because the minimum of the flare light curve goes down to the quiescent level.
Without adding some ad hoc new components to a simple hotspot model (like an obscuring component), it is clear that the flare light curve will always have a minimum above the quiescent level, since the hotspot will always be visible and make a non-zero contribution to the total intensity. 
As a consequence, we believe that fitting a hotspot-like model to our data would require some fine tuning using extra parameters, which would make the model less reliable.

\subsection{Constraining the radial distance of the first 2011 March 30 subflare from magnetic energy heating and synchrotron cooling}
We consider that the short duration of the rise phase of the first subflare places a limit to the size of the flaring region \citep[e.g.,][]{dodds-eden09}.
Following \cite{barriere14}, we assume that the energy released during the flare is powered during the rise phase by the magnetic energy available inside the flaring region, which constrains the radial distance of the flare.
By identifying the decay phase of the first subflare with synchrotron cooling, we derive a lower limit to the radial position of the first subflare.

\subsubsection{Timescales of the first subflare}
\begin{figure}[!h]
\centering
\includegraphics[trim = 0cm 0cm 0.55cm 0cm, clip,width=5.3cm,angle=90]{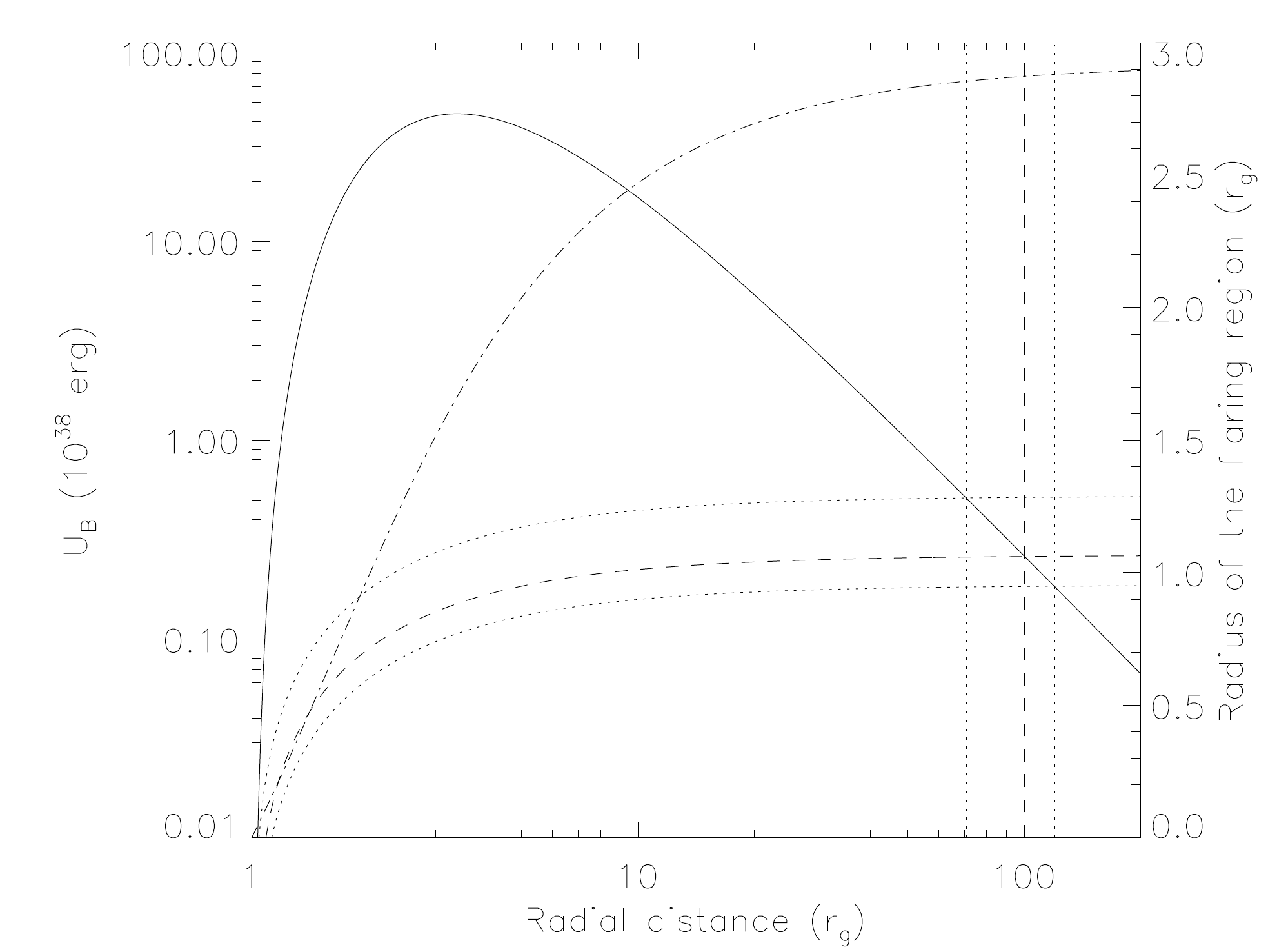}\\
\includegraphics[trim = 0cm 0cm 0.55cm 0cm, clip,width=5.3cm,angle=90]{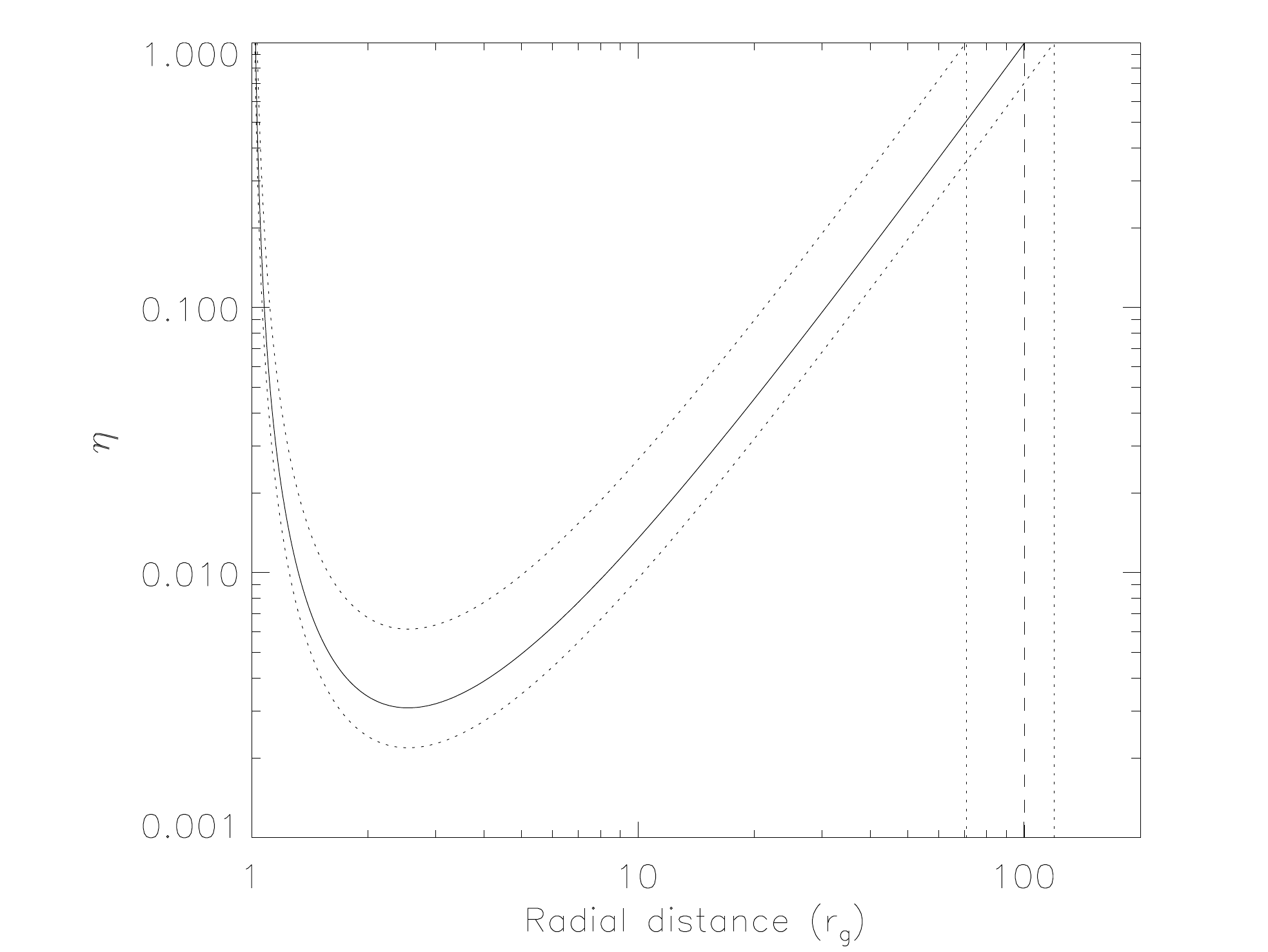}
\caption{Determination of the radial distance of the flaring region.
\textit{Top panel:} Magnetic energy vs. radial distance for a magnetic field of 100~G at $2\ r_{\mathrm{g}}$ and an X-ray photon production efficiency and dimensionless spin parameter of 1.
The solid line is the distribution of the magnetic energy (see left y-axis) vs. the radial distance.
The dashed and dotted lines represent the central value of the X-ray fluence and its errors with  90\% confidence level, respectively.
The vertical lines are the corresponding upper limits to the distance.
The dashed-dotted line represents the radius of the emitting region (see right y-axis).
\textit{Bottom panel:} X-ray photons production efficiency vs. radial distance for the fluence and its upper and lower limit.
The solid and dotted lines represent the efficiency for the central value of the fluence and its errors within 90\% confidence level, respectively.}
\label{Fig8}
\medskip
\includegraphics[trim = 0cm 0cm 0.5cm 0cm, clip,width=5.3cm,angle=90]{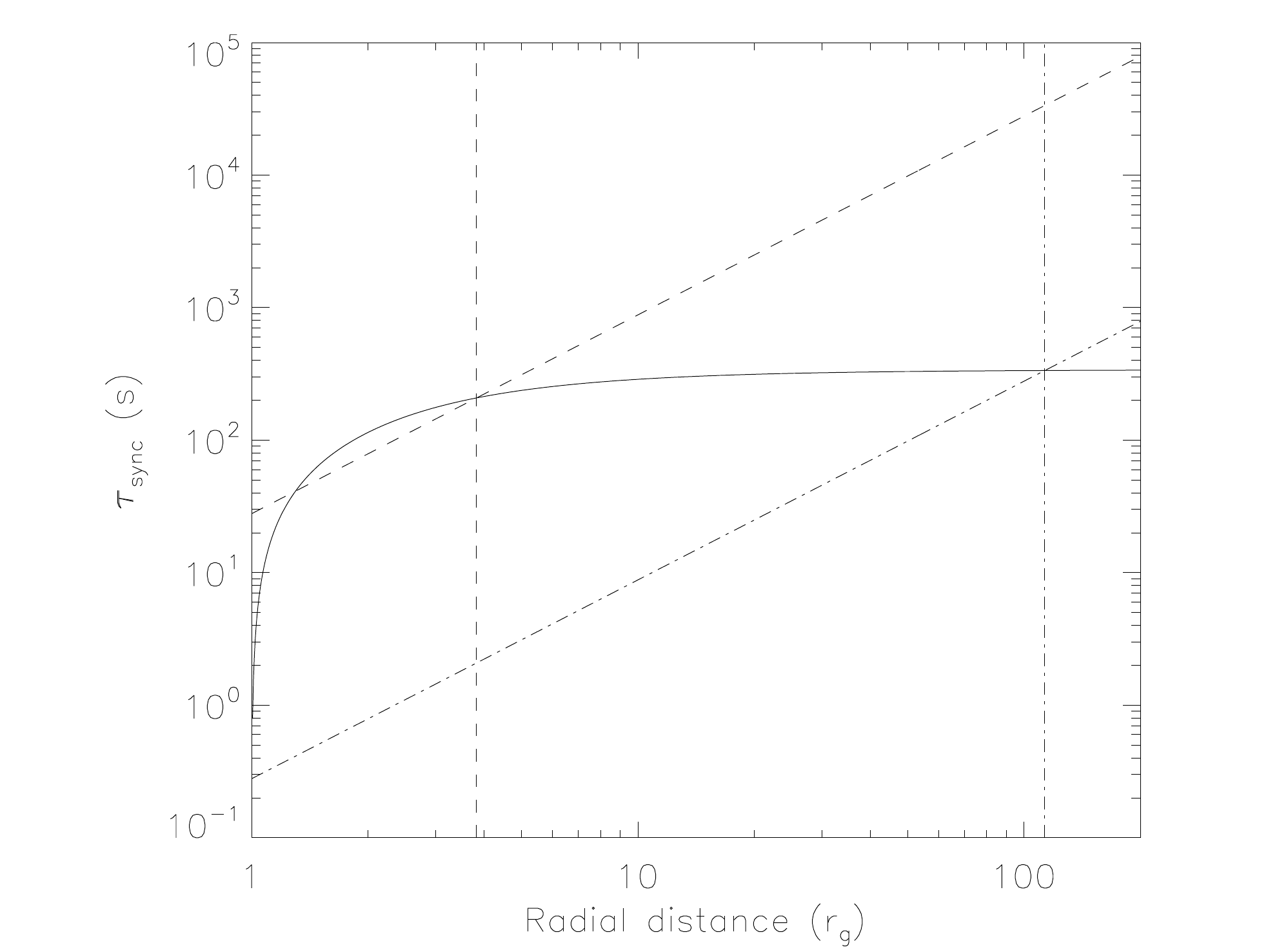}
\caption{ Synchrotron cooling time vs. the radial distance.
The solid line represents the proper duration of the decay phase.
The dashed inclined line represents the synchrotron cooling time for infrared photons.
The dotted-dashed inclined line is the synchrotron cooling time for X-ray photons.
The corresponding vertical lines are the lower limit to the radial distance for each cooling timescale.}
\label{Fig10}
\end{figure}
We define the start time of the rise phase as the time when the count rate of the smoothed light curve is too high to have been produced by the Poissonnian fluctuation of the non-flaring level at the 99.87\% of confidence level (corresponding to Gaussian single-sided confidence level of 3$\sigma$).
This threshold level is defined by $CR_0=N/h$, with $h$ the width of the kernel window ($h=100~\mathrm{s}$) and $N$ the lowest integer solution of the following equation:
\begin{equation}
 CDF= \sum_{n=0}^{N} \frac{(\lambda h)^n\,e^{-\lambda h}}{n!}  > 0.9987
\end{equation}
with $CDF$ the cumulative distribution function of the Poissonnian distribution, $\lambda$ the non-flaring rate (i.e., $0.107 \pm 0.001\ \mathrm{pn\ counts\ s^{-1}}$).
We find $CR_0=0.23\ \mathrm{pn\ counts\ s^{-1}}$ at $t_\mathrm{start}=17\mathrm{h}44\mathrm{m}59\mathrm{s}$ on 2011 March 30.
The end time of the rise phase is the maximum of the smoothed light curve that is reached at $t_\mathrm{max}=17\mathrm{h}46\mathrm{m}54\mathrm{s}$.
Thus, the rise phase duration is $\Delta t_{\mathrm{rise}}=t_\mathrm{max}-t_\mathrm{start}=115$ s.
The end time of the first subflare is the time of the minimum between the two subflares: $t_\mathrm{end}=17\mathrm{h}52\mathrm{m}34\mathrm{s}$.
This leads to $\Delta t_{\mathrm{flare}}=t_\mathrm{end}-t_\mathrm{start}=455$ s.

The proper duration of any event around an SMBH is always longer than the observed duration due to time dilation in strong gravity field. 
Therefore, we compute the proper-to-observed time ratio versus radial position (see Appendix \ref{appendix_c}). 
Hereafter we use a dimensionless spin parameter of one.

\subsubsection{Magnetic energy heating}
We constrain the radius of the spherical flaring region by considering that the Alfven velocity cannot be higher than the speed of light \citep{dodds-eden09}: $R<c \Delta\tau_{rise}$, where $\Delta \tau_{rise}$ is the proper duration of the rise phase.
This leads to the upper limit to the volume of the flaring region: $V=\frac{4}{3} \pi R^3<\frac{4}{3} \pi c^3 \Delta \tau^3_{\mathrm{rise}}$.
The magnetic energy contained inside this volume is $U_\mathrm{B} = \frac{B^2  V}{8 \pi}$ with $B=B_{1R_{\mathrm{S}}}2r_{\mathrm{g}}/r$ the magnetic field  vs. the radial distance $r$ \citep[see ][ and references therein]{barriere14}.

We define $\eta$, the X-ray photon production efficiency, as the ratio of the flare fluence in X-rays to the available magnetic energy.
The flare fluence in X-rays is the product of the unabsorbed X-ray luminosity with the duration of the first subflare (i.e., $\Delta \tau_{\mathrm{flare}}$).
Indeed, we have to compute the fluence released during the whole first subflare since all the X-ray emission from this event is powered by the magnetic heating of the emitting region.
We compute this luminosity with the parameters that were fitted to the flare spectrum, i.e., $N_\mathrm{H}=6.7 \times 10^{22}\ \mathrm{cm^{-2}}$ and $\Gamma=1.5$ (see first part of Sect. \ref{spectrum}).
The average luminosity of the first subflare is $\mathrm{L^{unabs}_{2-10~keV}(flare)}=5.8^{+5.7}_{-1.7} \times 10^{34}\ \mathrm{erg\ s^{-1}}$.
As a result, $\eta=\mathrm{L^{unabs}_{2-10~keV}(flare)} \Delta \tau_{\mathrm{flare}}/U_{\mathrm{B}}$.
Therefore, the upper limit to the radial distance can be computed by the relation
\begin{equation}
\mathrm{L^{unabs}_{2-10~keV}(flare)} \Delta \tau_{\mathrm{flare}}<\frac{B^2_{1R_{\mathrm{S}}}}{6} \left(\frac{2r_{\mathrm{g}}}{r}\right)^2 c^3  \Delta \tau^3_{\mathrm{rise}} \eta \, .
\end{equation}
If we assume a maximum efficiency ($\eta=1$), the upper limit to the radial distance is $r<100^{+19}_{-29}\ r_{\mathrm{g}}$ (see Fig. \ref{Fig8}).
The corresponding radius of the flaring region at this distance is $R=2.87\pm 0.01\ r_{\mathrm{g}}$.

We can neglect any magnification of the observed luminosity compared to the proper luminosity at this radial distance.
Indeed in the hotspot model, the combined effects of the beaming and the gravitational redshift on the proper luminosity are small at $r=100\ r_{\mathrm{g}}$ since the corresponding orbital period is $\sim$1.5~days, which implies that any magnification has a long timescale and a very small amplitude \citep{broderick05,hamaus09}.
In the jet geometry, the Doppler factor is small owing to the small inclination and the mild velocity of the Sgr~A* jet; therefore, the beaming factor, varying as the square of the Doppler factor, is also small \citep{barriere14}.

\subsubsection{Synchrotron cooling}
The electrons that were accelerated by the release of the magnetic energy will cool by emitting synchrotron radiation with the following timescale:
$\tau_\mathrm{sync}=8 \times \left( B/30~\mathrm{G}\right)^{-3/2} \times \left( \nu/10^{14}~\mathrm{Hz}\right)^{-1/2}$~min 
\citep{dodds-eden09}.
If the X-ray photons at $10^{18}$ Hz are the primary source of synchrotron cooling, then $\tau_\mathrm{sync}^\mathrm{X}=0.78 \left(B_{1R_{\mathrm{S}}}/100~\mathrm{G} \right) \left(r/2r_{\mathrm{g}}\right)^{3/2}$~s. 
From $\tau_\mathrm{sync}^\mathrm{X}>\Delta \tau_\mathrm{decay}$, we derive $r>114~r_{\mathrm{g}}$, which is not consistent with the previously derived upper limit. 
Therefore, if the X-rays are the primary source of synchrotron cooling in this subflare, sustained heating must also be present during the decay phase.

We know that X-ray flares are always associated with NIR flares \citep[e.g.,][]{dodds-eden09}, which have power-law spectra consistent with synchrotron process \citep{eisenhauer05}.
Thus, we consider the synchrotron cooling time of NIR photons ($\nu=10^{14}$ Hz)--$\tau_\mathrm{sync}^\mathrm{NIR}=78.9 \left(B_{1R_{\mathrm{S}}}/100~\mathrm{G} \right) \left(r/2r_{\mathrm{g}}\right)^{3/2}$~s--which leads to $r>4\ r_{\mathrm{g}}$ with the flaring region outside the event horizon.
The evolution of these synchrotron cooling times with the radial distance is shown in Fig. \ref{Fig10}.

We conclude that $4\ r_{\mathrm{g}}<r<100^{+19}_{-29}\ r_{\mathrm{g}}$ in this subflare for $\eta=1$ and $B_{1R_{\mathrm{S}}}=100~\mathrm{G}$. 
The corresponding radii of the flaring region at these distances are $1.8\ r_{\mathrm{g}}<R<2.87\pm 0.01\ r_{\mathrm{g}}$.
The minimum distance of $r>1.9\ r_{\mathrm{g}}$ is required to have the flaring region well outside the event of horizon.

\subsubsection{Comparison with previous works}
The upper limit to the radial distance of the first subflare on 2011 March 30 is five times more than the one derived for the flare detected by NuSTAR on 2012 July 21 \citep[][]{barriere14}. 
The latter was longer (1896~s) and about $\sim3.5$ times more luminous (mean luminosity of $21 \times 10^{34}\ \mathrm{erg\ s^{-1}}$) than the former.
Moreover, the 2012 July 21 NuSTAR flare was characterized by a plateau phase of $\approx$1700~s between the rise and decay phases of 100~s.

\cite{barriere14} assume that the radius of the emitting region is constant after the rise phase. 
But for a radial position lower than $20\ r_{\mathrm{g}}$ and $B_{1R_{\mathrm{S}}}=100$~G, the synchrotron cooling time of NIR photons is lower than 2500~s, which implies that the heating process is still required after the rise phase to produce the observed plateau phase. 
Therefore, the radius of the emitting region of this NuSTAR flare is likely much larger than those computed by \cite{barriere14} from only the rise phase.
If we remove the plateau phase of this NuSTAR flare, only the rise and decay phases remain, so, we can use the same  method as for our first subflare.

We revise the upper limit to the radial distance of the 2012 July 21 NuSTAR flare to at least $64\ r_{\mathrm{g}}$.
Including the likely increase in the radius of the flaring region during the plateau phase due to sustained heating leads to an even higher value for the upper limit to the radial distance.

\section{Summary}\
\label{summary}
We have reported the data analysis of the XMM-Newton 2011 campaign observation of \sgra{} (five observations with a total of exposure of $\approx 226\ \mathrm{ks}$).
We used the Bayesian-blocks algorithm developed by \citet{scargle98} and a density estimator with an Epanechnikov kernel to constrain the duration, the position, and the amplitude of the X-ray flares with better accuracy.
The Bayesian-blocks algorithm uses the unbinned event arrival time on the EPIC cameras to identify the flaring and non-flaring period and their corresponding count-rate levels.
This analysis of the event's arrival time increases the accuracy on the time of the beginning and the end of a flare in comparison with a detection above a given threshold of a binned light curve.
The algorithm uses a Bayesian statistic to find the time when  the count-rate level is statistically different under a given probability.
We worked with a false detection probability of $\exp(-3.5),$ which implies that the detected flare is a real flare with a probability of 99.9\%.
We corrected the contribution of the flaring background by applying twice this algorithm on the source and the background regions.
We used a density estimator to improve the determination of the characteristics of the flares.
The density estimator applies a convolution between the event list corrected from the GTI and a kernel defined on a finite support in order to control any boundary effects.
Thanks to the Bayesian-blocks algorithm, we could also correct the resulting smoothed light curves from the flaring background.

We observed two X-ray flares during these observations.
The former occurred on 2011 March 30 and the latter on 2011 April 03.
For comparison, these flares have a peak detection level of 6.8 and 5.9 $\sigma$ in the XMM-Newton/EPIC (pn+MOS1+MOS2) light curve in the 2$-$10~keV energy range with a 300 s bin.
The first flare is composed of two subflares: a very short-duration ($\sim 458$ s) one with a peak luminosity of $L\mathrm{^{unabs}_{2-10~keV}} \sim 9.4 \times 10^{34}\ \mathrm{erg\ s^{-1}}$ and a longer ($\sim 1542$ s) and less luminous one ($L\mathrm{^{unabs}_{2-10~keV}} \sim 6.8 \times 10^{34}\ \mathrm{erg\ s^{-1}}$ at the peak).
The spectral analysis of this flare allowed us to derive these parameters:  $N_\mathrm{H}=6.7^{+8.2}_{-6.7} \times 10^{22}\ \mathrm{cm^{-2}}$, $\Gamma=1.5^{+1.5}_{-1.3}$, $F\mathrm{^{abs}_{2-8~keV}}=2.5 \times 10^{-12}\ \mathrm{erg\ s^{-1}\ cm^{-2}}$, and $F\mathrm{^{unabs}_{2-10~keV}}=3.5^{+3.1}_{-1.0} \times 10^{-12}\ \mathrm{erg\ s^{-1}\ cm^{-2}}$.
These spectral parameters are consistent with those previously found by \citet{porquet03,porquet08} and \citet{nowak12} but are not really constrained.

A comparison of the physical characteristics of this flare with those reported by \citet{neilsen13} from the 2012 \textit{Chandra XVP} campaign shows that it lies in the mean of the X-ray flares detected by Chandra, but the first subflare is one of the shortest and less luminous X-ray flares.
The distribution of the minimum waiting time between two successive flares in the \textit{Chandra XVP} campaign favors the hypothesis of a single flare.

We modeled its two subflares with a single physical phenomenon using the gravitational lensing of a hotspot-like structure.
However, the consistency of the flux level between the two subflare peaks with the non-flaring one led us to conclude that the light curve of this X-ray flare cannot satisfactorily be reproduced by a gravitational lensing event.

We also constrained the radial position of the emitting region of the first 2011 March 30 subflare by assuming that the heating energy is provided by the magnetic field available in the spherical emitting region whose radius is determined by the duration of the rise phase of this first subflare.
A comparison of the duration of the decay phase of this subflare and the synchrotron cooling timescale allowed us to determine a lower limit to the radial distance.
We conclude that the X-ray emitting region of the first subflare is located at a radial position of $4-100^{+19}_{-29}$ and has a corresponding radius of $1.8- 2.87 \pm 0.01$ in $r_{\mathrm{g}}$ unit for a magnetic field of 100 G at $2\ r_{\mathrm{g}}$.

\begin{acknowledgements}
E.M. acknowledges the Université de Strasbourg for her IdEx PhD grant.
The XMM-Newton project is an ESA Science Mission with instruments and contributions directly funded by ESA Member States and the USA (NASA).
This work has been financially supported by the Programme National Hautes Energies (PNHE).
The research leading to these results received funding from the European Union Seventh Framework Program (FP7/2007-2013) under grant agreement n$^{\circ}$312789.
\end{acknowledgements}

\bibliographystyle{aa}
\bibliography{biblio_centre_galactique.bib}
\Online

\appendix

\section{Calibration of the ncp\_prior relation}
\label{appendix_a}
\begin{figure}[!h]
\centering
\includegraphics[trim = 9cm 2cm 3cm 2cm, clip, angle=180,width=9.6cm]{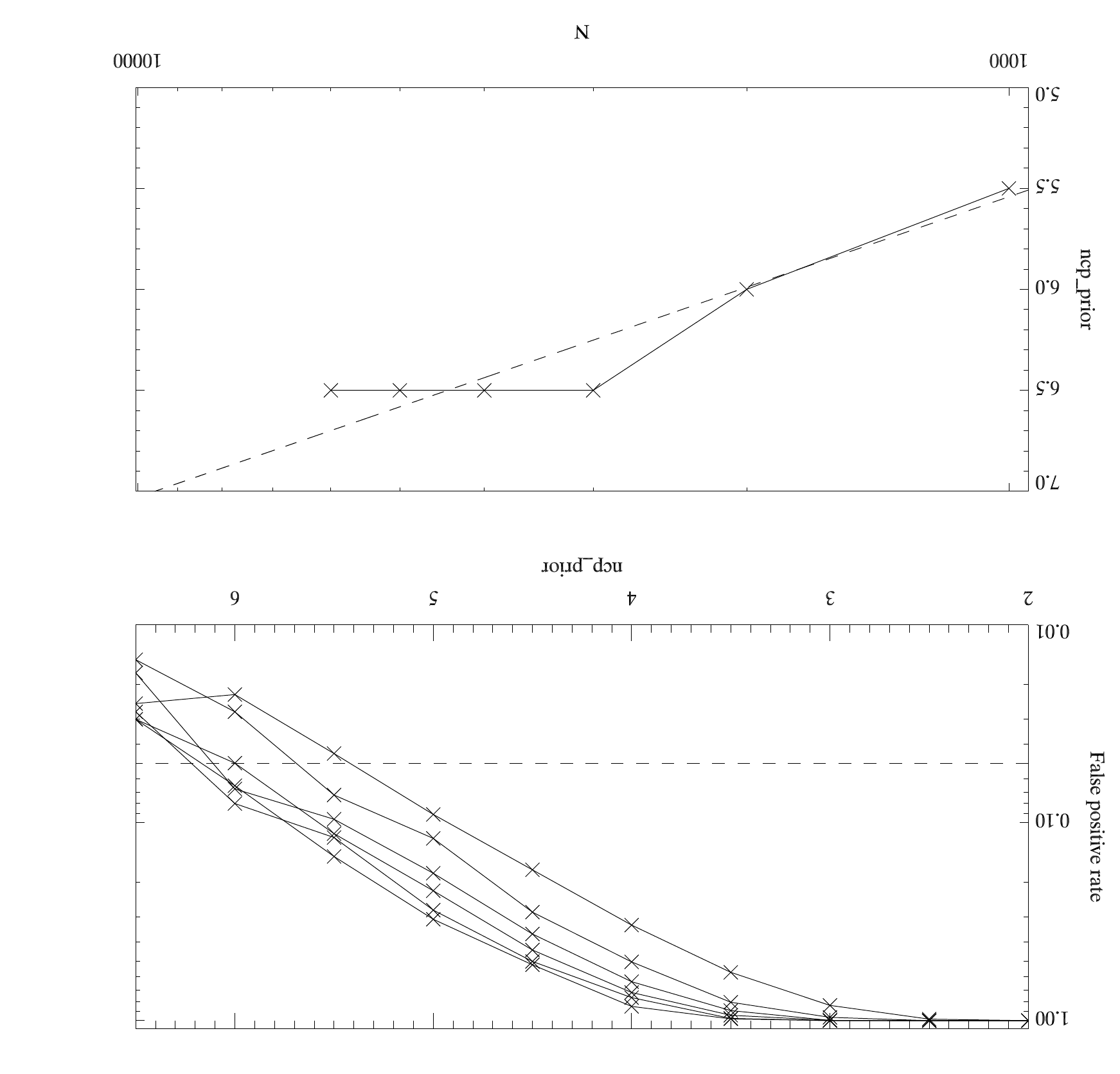}
\caption{Simulations of point measurements (Poisson signal of average 0.1) to determine $ncp\_ prior = - log(\gamma)$.
\textit{Top:} false positive fraction $p_\mathrm{1}$ vs. value of $ncp\_ prior$ with separate curves for the values $N = 1000, 2000, 3000, 4000, 5000$, and $6000$ (left to right). 
The points at which the rate becomes unacceptable (here 0.05; dashed line) determine the recommended values of $ncp\_prior$ shown as a function of $N$ in the bottom panel.
\textit{Bottom:} Calibration of $ncp\_prior$ as a function of the number of counts (N) for a value of $p_\mathrm{1}$ (here: 0.05).
The dashed line is the linear fit of the simulation points.}
\label{Fig4}
\end{figure}

We cannot use the scaling relation given in \citet{scargle13b} for our data set because it has different statistical properties than the simulated data set used by \citet{scargle13b}.
First, our events are affected by Poissonian noise and not by Gaussian noise.
Second, our event lists with about $\sim 4000$ counts is longer than the published simulation limited to 1024 counts.
To calibrate the relation between $ncp\_prior$ (the prior of the number of block) and the false positive rate ($p_\mathrm{1}$), we simulate 100 constant light curves with Poisson noise around a level of $0.1\ \mathrm{count}\ \mathrm{s^{-1}}$, which is the typical non-flaring level measured by XMM-Newton/EPIC pn during our observations.
For each sequence of 100 simulations, we increase the $ncp\_prior$ value from 2 to 9 by a step of 0.5 and we compute the number of change points detected. 
The percentage of change points detected in 100 simulations determines the $p_\mathrm{1}$. 
We repeat this operation for different numbers of count $N$ in the light curve (from 1000 counts to 6000 counts by step of 1000 counts).
With the $p_\mathrm{1}$ values and the corresponding $ncp\_prior$, we can create the graph presented in Fig. \ref{Fig4}.
Then, we can take different values of $p_\mathrm{1}$ and report the relation between the count number and $ncp\_prior$ that satisfied $p_\mathrm{1}$. 
An example with $p_\mathrm{1}=0.05$ is given in the bottom graph of Fig. \ref{Fig4}. The dashed line is the linear fit of the curve.
Thus, we have the same number of relations between $N$ and $ncp\_prior$ as the number of value of $p_\mathrm{1}$ that we choose.
By combining these relations, which relies $p_\mathrm{1}$, $N,$ and $ncp\_prior$, we find our calibration: 
\begin{equation}
   \begin{array}{ll}
  ncp\_prior = & 3.356+0.143 \ln \left(N\right)-0.710 \ln \left(p_\mathrm{1}\right)\\
   & -0.002 \ln \left(N\right) \ln \left(p_\mathrm{1}\right)
    \end{array}
\end{equation}
with $N$ the number of events in a range of [1000:6000] counts.
For $N$ lower than 1000, the last term is lower than 0.01, which is negligible.
For a probability of false detection equals $\exp(-3.5)$ and $\mathrm{N}=4000$, $\mathrm{ncp\_prior}=7.0099$.

\section{Detection rate vs. flare peak and duration}
\label{appendix_b}
\begin{figure}[!h]
\centering
\includegraphics[trim = 15cm 9.65cm 3cm 3.3795cm, clip, angle=180,width=7cm]{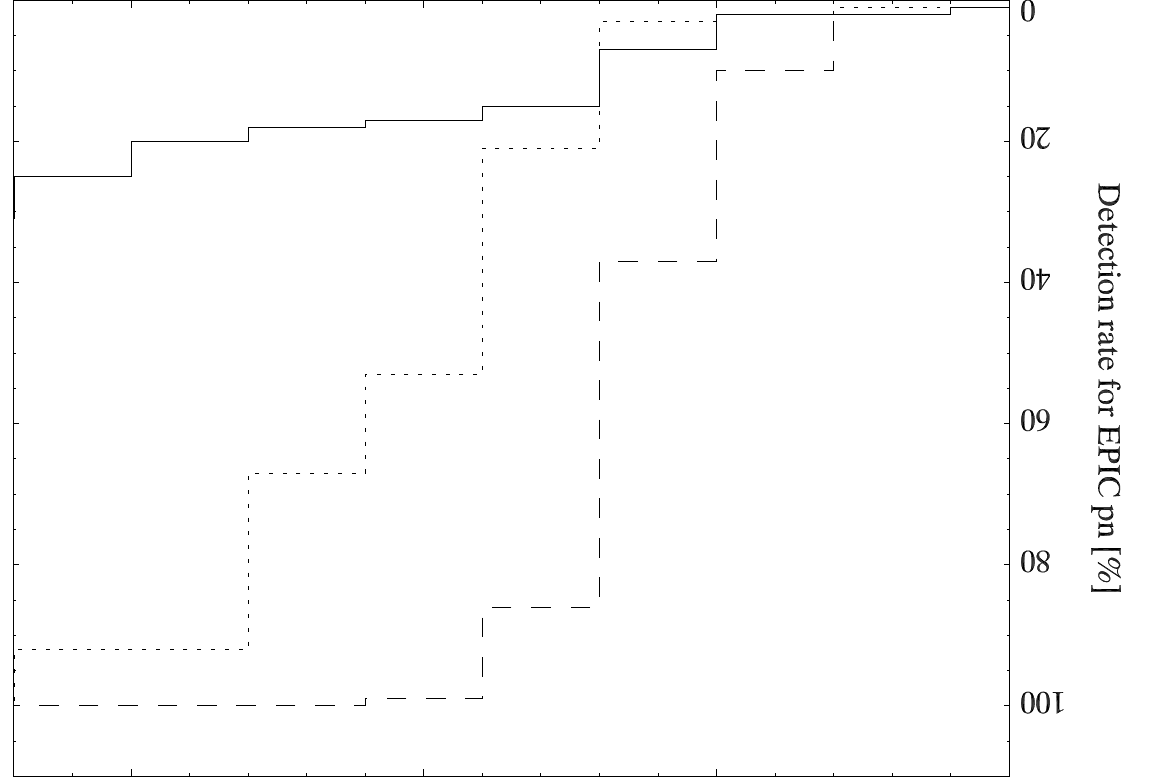}\\
\includegraphics[trim = 2cm 9.741cm 16cm 2cm, clip, angle=180,width=7cm]{24682_figB1.pdf}
\caption{Detection level for different values of Gaussian amplitude and $p_\mathrm{1}=\exp(-3.5)$. 
The solid line corresponds to $FWHM=56.62\ \mathrm{s}$, the dotted line corresponds to $FWHM=318.49\ \mathrm{s,}$ and the dashed line corresponds to $FWHM=1104\ \mathrm{s}$. }
\label{Fig5}
\end{figure}

To evaluate our detection level, we simulate light curves with a Poisson signal of average $0.1\ \mathrm{count}\ \mathrm{s^{-1}}$ for EPIC pn and $0.04\ \mathrm{count}\ \mathrm{s^{-1}}$ for EPIC MOS corresponding to the non-flaring level of these cameras.
This difference in the non-flaring level between the two cameras implies a difference in the Poisson noise (the higher the non-flaring level, the higher the Poisson noise), hence in the detection rate.
On these constant light curves, we add a Gaussian with a FWHM equal to $1104\ \mathrm{s}$, $318.49\ \mathrm{s,}$ and $56.62\ \mathrm{s,}$ which correspond to the maximum, the median, and the minimum, respectively, of the FWHM of the X-ray flares from \sgra{} detected by Chandra and reported by \citet{neilsen13}. 
We vary the amplitude of the Gaussian between $0$ and $0.17\ \mathrm{count}\ \mathrm{s^{-1}}$ above the non-flaring level.
For each amplitude, we perform 100 simulations and compute the number of flare (two change points) found by the Bayesian-blocks algorithm for a false positive rate equal to $\exp(-3.5)$. 
The results are shown in Fig. \ref{Fig5}.

\begin{figure}[!t]
\centering
\includegraphics[angle=90,width=5.9cm]{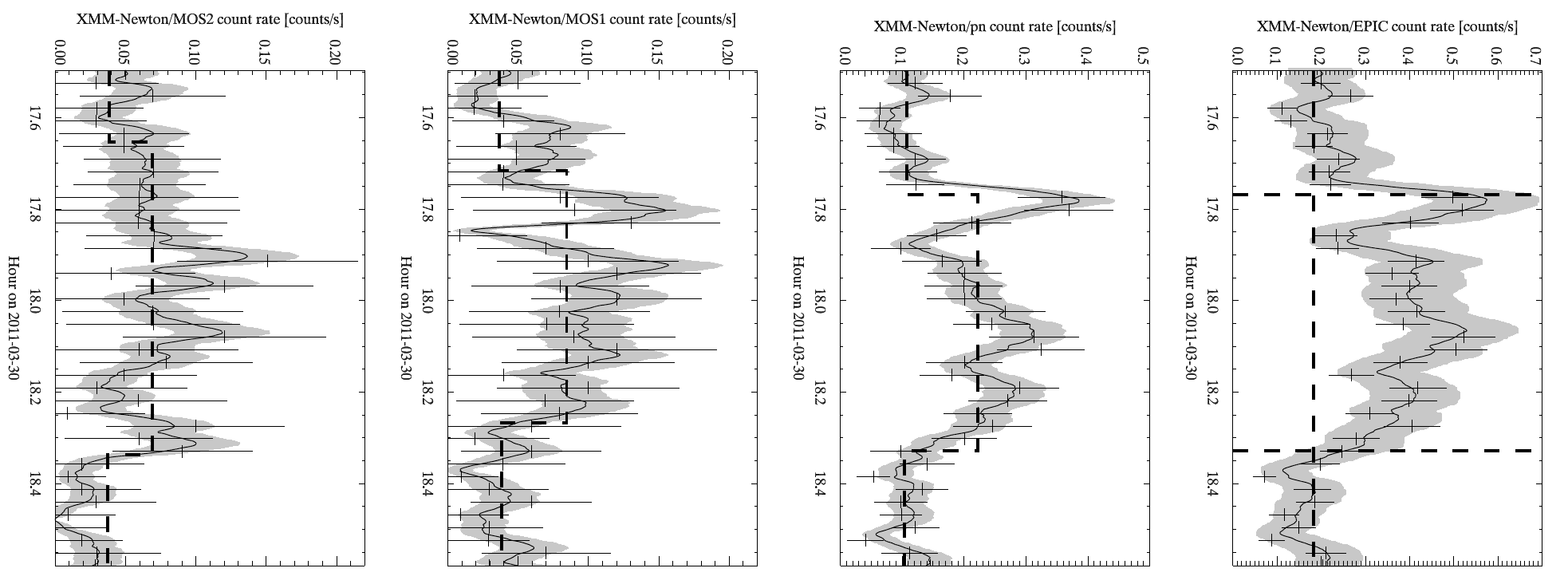}\\
\caption{Light curves of \sgra{} in the 2–10 keV energy range obtained during the flare on 2011 March 30 binned on 100s.
\textit{Top:} The total XMM-Newton/EPIC light curve.
The horizontal dashed line represents the non-flaring level calculated as the sum of the non-flaring level in each instrument found by the Bayesian blocks.
The vertical dashed lines represent the beginning and the end of the flare calculated by the Bayesian-blocks algorithm on pn camera.
The solid line is the smoothed light curve that is the sum of the smoothed light curve for each instrument (calculated on the same time range).
The gray curve shows the errors associated with the smoothed light curve.
In all panels, the time period during which the camera did not observe is shown by a light gray box.
\textit{Second panel:} The XMM-Newton/EPIC pn light curve of \sgra.
\textit{Third panel:} The XMM-Newton/EPIC MOS1 light curve of \sgra.
The vertical dashed lines represent the beginning and the end of the flare calculated by the Bayesian-blocks algorithm on MOS1 camera.
\textit{Bottom panel:} The XMM-Newton/EPIC MOS2 light curve of \sgra.
The vertical dashed lines represent the beginning and the end of the flare calculated by the Bayesian-blocks algorithm on MOS2 camera.}
\label{Fig6}
\end{figure}
\begin{figure}[!Ht]
\centering
\includegraphics[angle=90,width=5.9cm]{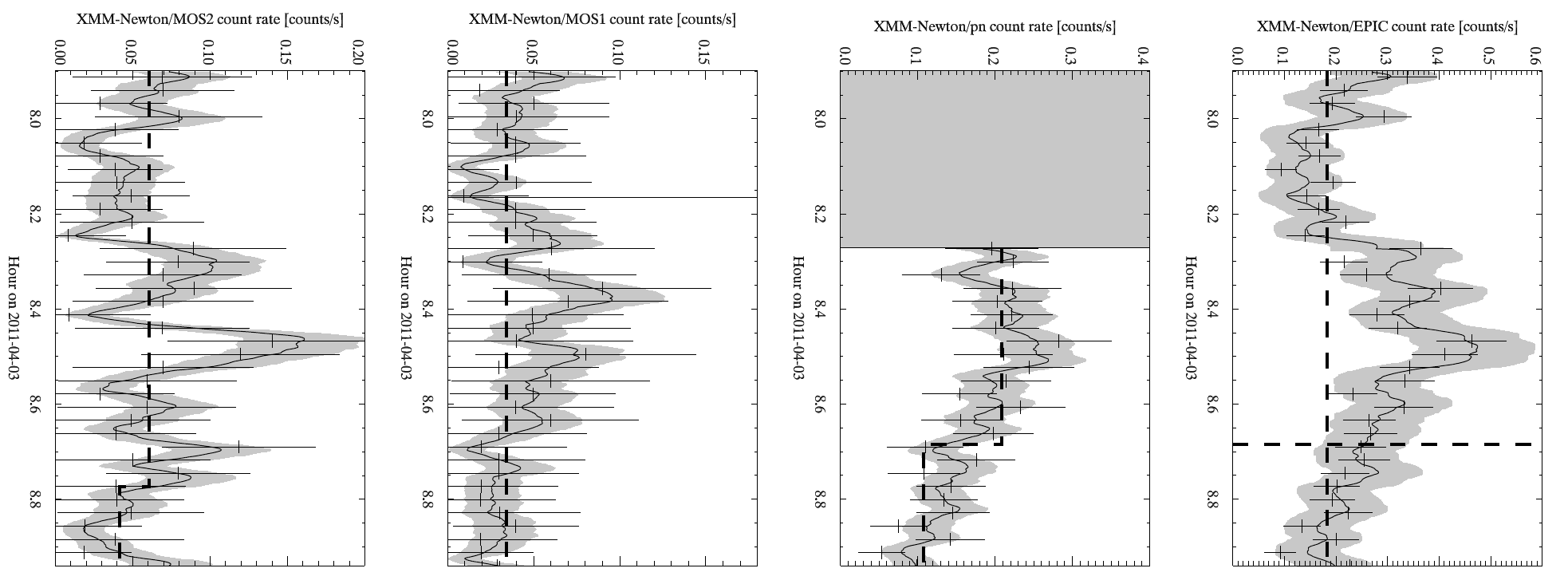}
\caption{Light curves of \sgra{} in the 2–10 keV energy range obtained during the flare on 2011 April 3 binned on 100s.
\textit{Top:} The total XMM-Newton/EPIC light curve.
The horizontal dashed line represents the non-flaring level calculated as the sum of the non-flaring level in each instrument found by the Bayesian blocks.
The vertical dashed lines represent the beginning and the end of the flare calculated by the Bayesian-blocks algorithm on pn camera.
The solid line is the smoothed light curve that is the sum of the smoothed light curve for each instrument (calculated on the same time range).
The gray curve shows the errors associated with the smoothed light curve.
In all panels, the time period during which the camera did not observe is shown by a light gray box delimited by vertical solid lines.
\textit{Second panel:} The XMM-Newton/EPIC pn light curve of \sgra.
The dark gray box is the time during which pn did not observe.
\textit{Third panel:} The XMM-Newton/EPIC MOS1 light curve of \sgra. 
The light gray vertical line shows the time during which MOS1 did not observe.
The vertical dashed lines represent the beginning and the end of the flare calculated by the Bayesian-blocks algorithm on MOS1 camera.
\textit{Bottom panel:} The XMM-Newton/EPIC MOS2 light curve of \sgra.
The vertical dashed lines represent the beginning and the end of the flare calculated by the Bayesian-blocks algorithm on MOS2 camera.}
\label{Fig7}
\end{figure}
We can see that the higher the amplitude and the FWHM of the flare, the higher the detection rate.
We can also see that the main difference between the detection rate in the XMM-Newton/EPIC MOS and pn camera (the former has a non-flaring level that is two times lower than in pn) is that the small flares with large FWHM are more detected in MOS than in pn.

Figures \ref{Fig6} and \ref{Fig7} show the flare light curves obtained with of XMM-Newton/EPIC observed on 2011 March 30 and April 3.
We can see that the first and second subflares on 2011 March 30 are distinguishable on XMM-Newton/EPIC pn and MOS1 but not in MOS2 even if a flare is detected by the Bayesian-blocks algorithm.
The flare on 2011 April 3 is not detected by the Bayesian-blocks algorithm in XMM-Newton/EPIC MOS1.
This is because the algorithm allows us to find a flare whose $FWHM \approx 900\ \mathrm{s}$ in EPIC MOS camera with a probability of $95$\% if its amplitude above the non-flaring level is higher than $0.07\ \mathrm{count}\ \mathrm{s^{-1}}$ with a probability of false detection equal to $\exp(-3.5),$ but in XMM-Newton/EPIC MOS1, the flare amplitude is about $0.06\ \mathrm{count}\ \mathrm{s^{-1}}$.
Since XMM-Newton/EPIC MOS1 and MOS2 have lower number counts than XMM-Newton/EPIC pn because of the RGS, it is on XMM-Newton/EPIC pn that the flare will have higher amplitude and thus higher accuracy on the determination of the beginning and end of the flare.

\section{Time dilatation around a Kerr black hole}

\label{appendix_c}
We use the Kerr metric in Boyer-Lindquist coordinates:
\begin{equation}
   \begin{array}{ll}
           ds^2=-d\tau^2= & -\left(1- \frac{2r}{\Sigma}\right)dt^2-\frac{4ar\sin^2\!\theta}{\Sigma}dt\,d\phi + \frac{\Sigma}{\Delta}dr^2 \\
      & + \Sigma\, d\theta^2+ \left(r^2+a^2+\frac{2a^2r\sin^2\!\theta}{\Sigma}\right)\sin^2\!\theta\, d\phi^2
    \end{array}
\end{equation}
with $\tau$ the proper time, $t$ the observed time, $r$ the radial distance in gravitational radius, $a$ the dimensionless spin parameter, $\Sigma=r^2+a^2\,\cos^2\!\theta$, $\Delta=r^2-r+a^2$, and $\theta=0$ defining the spin axis \citep{bardeen72}.
For a direct circular orbit in the equatorial plane, we have $\frac{dr}{dt}=0$, $\theta=\frac{\pi}{2}$, and $\frac{d\phi}{dt}=\frac{1}{r^{3/2}+a}$ \citep{bardeen72}.
\setcounter{section}{3}
\setcounter{figure}{0}
\begin{figure}[!h]
\centering
\includegraphics[angle=90,width=7cm]{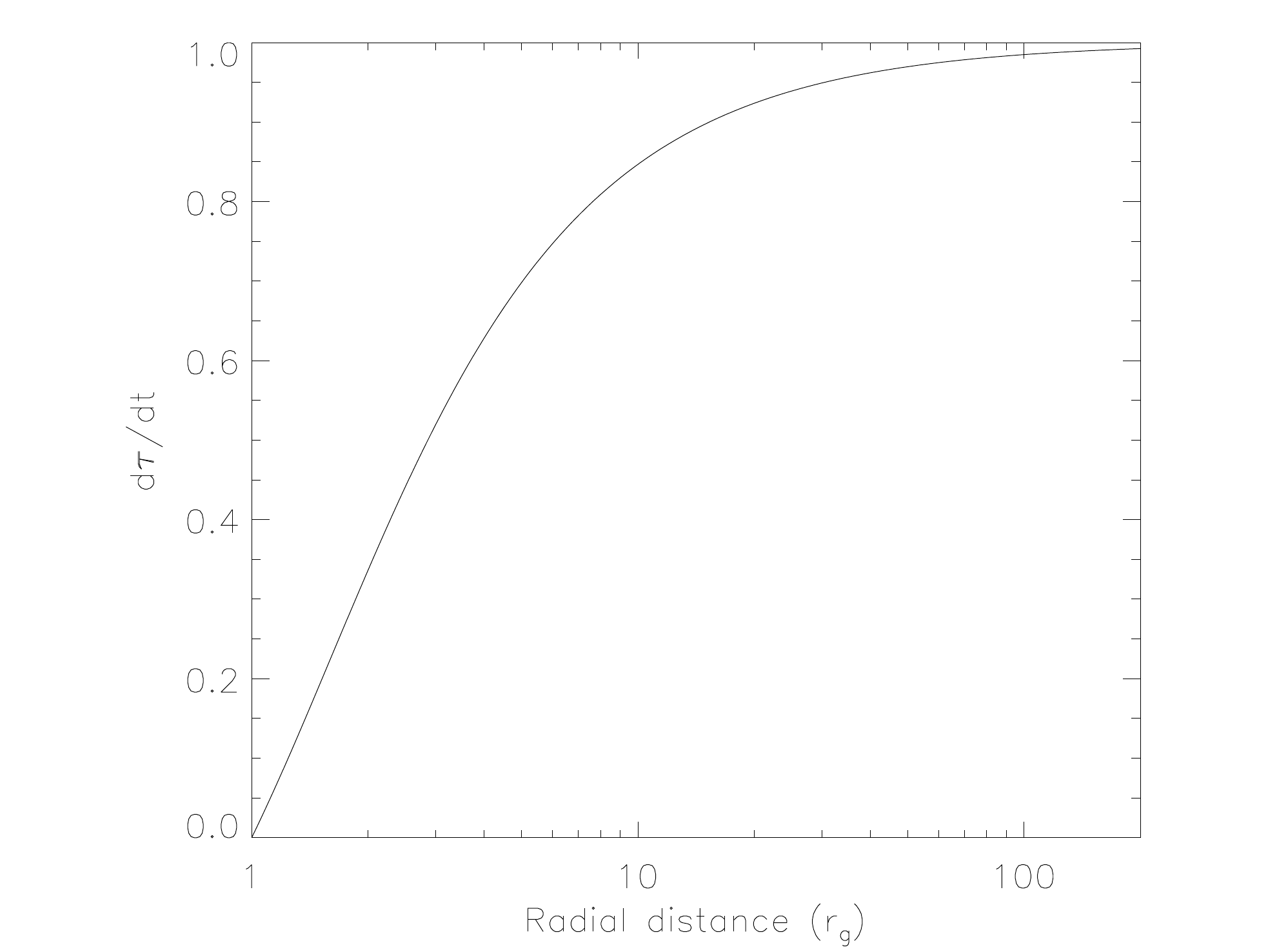}
\caption{ Ratio between the proper time and the observed time close to a Kerr black hole with a dimensionless spin parameter of 1. }
\label{Fig9}
\end{figure}
\setcounter{section}{2}
\setcounter{figure}{1}
Thus, the relation between the proper time and the observed time is 
\begin{equation}
\frac{d\tau}{dt}=\sqrt{1-\frac{2}{r} - \frac{r^3-4ar^{3/2} + a^2r-2a^2}{r\left(r^{3/2} + a \right)^2}}\end{equation}
Figure \ref{Fig9} shows the time dilatation as a function of the radial distance plotted from the innermost boundary of the circular orbit, i.e., $r_\mathrm{g}$ for $a=1$.


\newpage
\includepdf{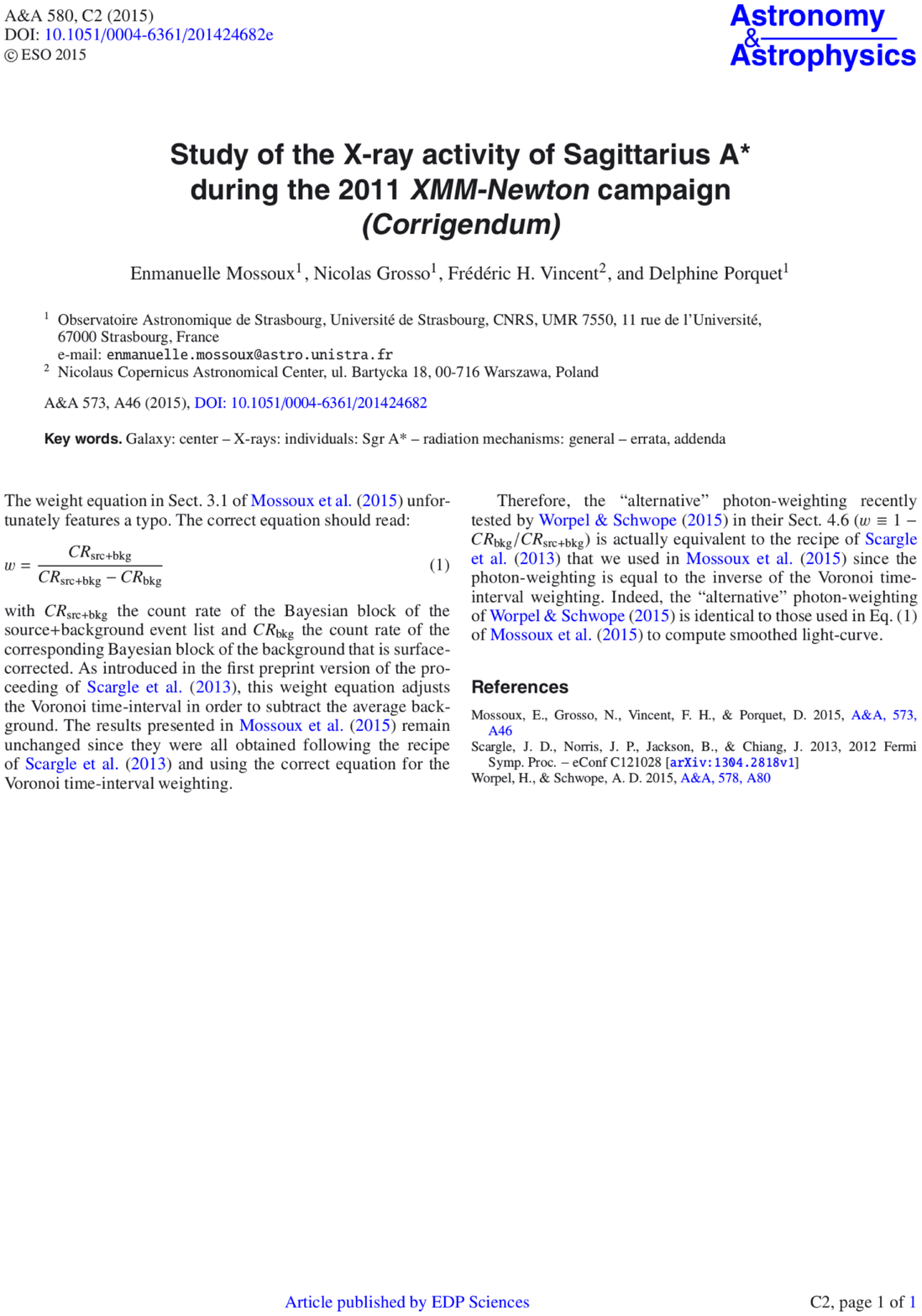}
\end{document}